\documentclass[aps,pra,reprint,superscriptaddress,amsmath,amssymb,showkeys,floatfix]{revtex4-1}

\usepackage{xcolor}

\DeclareMathOperator{\tr}{tr}

\usepackage{tabularx}
\usepackage{xspace}

\usepackage{graphicx} 
\usepackage{float}
\usepackage{bbold}
\usepackage{bm}
\usepackage{amsthm}
\newtheorem{theorem}{Theorem}
\newtheorem{result}[theorem]{Result}
\usepackage[bookmarks=false,pdfstartview={FitH}]{hyperref}

\begin{document}
\title{Wigner process tomography: Visualization of spin propagators and their spinor properties}
\author{David Leiner}
\email[]{david.leiner@tum.de}
\author{Steffen J. Glaser}
\email[]{steffen.glaser@tum.de}
\affiliation{Technische Universit\"at M\"unchen, Department Chemie, Lichtenbergstrasse 4, 
85747 Garching, Germany}
\date{\today}

\begin{abstract}
We study the tomography of propagators for spin systems
in the context of finite-dimensional Wigner representations,
which completely characterize and visualize operators
using shapes assembled from linear combinations of spherical harmonics.
The Wigner representation of a propagator can be experimentally recovered by
 measuring expectation values of rotated 
axial spherical tensor operators in an augmented system with an additional ancilla qubit. The methodology is experimentally demonstrated for standard one-qubit quantum gates using nuclear magnetic resonance spectroscopy. In particular, this approach provides a direct and compelling visualization of the spinor property of the propagators corresponding to the rotation of a spin 1/2 particle.
\end{abstract}

\maketitle

\section{Introduction}
Phase-space representations provide useful tools for the characterization and visualization of quantum systems \cite{SchleichBook, Curtright-review,deGossonbook}. 
Here we consider continuous Wigner representations of individual \cite{Stratonovich, DowlingAgarwalSchleich} and coupled \cite{JHKS,Garon15, tilma2016} spin systems. In particular, we focus on the so-called DROPS (discrete representation of operators for spin systems) representation \cite{Garon15, DavidTomo}, which provides an intuitive visualization of the states and operators of coupled spin systems, reflecting physically relevant properties, such as
symmetries with respect to rotations and permutations of spins. The DROPS representation has also been implemented in a free, interactive application software \cite{ipad_app,glaserapp20}.
Following the general strategy of Stratonovich \cite{Stratonovich},
which specifies criteria for the definition of
continuous Wigner functions for finite-dimensional quantum systems,
the DROPS representation is based on a mapping of arbitrary operators to a set of spherical functions
which are denoted as {\it{droplets}}.
In particular, as illustrated in \cite{Garon15,Damme}, the DROPS representation is also applicable to propagators and
not limited to density operators.

We recently studied an experimental quantum state tomography scheme to scan 
 generalized Wigner representations of 
the {\it density operator} for arbitrary multi-spin quantum states \cite{DavidTomo}. 
We also provided explicit experimental protocols for our Wigner tomography scheme
and demonstrated its feasibility using 
nuclear magnetic resonance (NMR) experiments.

In contrast to {\it state tomography}, where the purpose is to characterize the state of a system,
the aim of {\it process tomography} \cite{Chuang1997,Poyatos1997,Schmiegelow2011,PhysRevA.64.012314,Altepeter2003,Gaikwad2018} 
is to fully characterize a quantum process that 
can be applied to arbitrary states.
In the standard process tomography
scheme, a set of defined input states is prepared and the output of the unknown quantum process is measured.
In general, a
quantum process can be described by a completely positive map \cite{Heinosaari2011}.
In the special case of closed quantum systems with negligible relaxation, 
quantum processes are characterized by a unitary time evolution operator, the so-called {\it propagator}.

In this work we ask wether our earlier approach \cite{DavidTomo} for the tomography of the Wigner representation of quantum states can be extended to experimentally scan 
the Wigner representation of {\it propagators}. 
This would lead to an 
alternative form of (Wigner) process tomography for the characterization of pulse 
sequence elements or entire pulse sequences in spectroscopy and in quantum information 
processing (quantum gates, quantum algorithms).
The idea is to imprint a given propagator onto the density operator $\rho$ and to use a variant 
of the scanning scheme introduced in \cite{DavidTomo} to reconstruct the Wigner representation of 
these propagators. We will focus on systems consisting of spins 1/2, even though our approach is 
applicable to arbitrary spin systems.
Additionally, explicit experimental protocols for our Wigner tomography scheme are provided and  experimentally demonstrated using methods of NMR.

The proposed Wigner tomography of propagators also  provides a direct way to visualize the spinor properties of spin-1/2 rotation operators. Following
\cite{Thompson1994}, a spinor is defined as a mathematical entity that changes its sign under a 2$\pi$ rotation. As discussed in \cite{Thompson1994},
the {\it propagators} for the rotations of half-integer spins are spinors and a direct consequence 
of this property is that also the {\it state vectors} of half-integer spins are spinors. 
Previous works regarding the measurement of the spinor property of state vectors 
were based on neutron interferometry~\cite{Werner,Badurek,Mezei,Massimiliano} and also NMR spectroscopy~\cite{Stoll,Suter,Mehring}. 
However, to our knowledge the underlying spinor property of entire {\it propagators} has so far not explicitly been demonstrated.
As shown in the following, this property can be directly visualized by observing the sign change of the Wigner representation of rotation propagators.

The paper is organized as follows.  An overview of the representation and visualization method for coupled spin systems is given in Sec. \ref{chapt:summary_drops}. A brief summary of the 
scanning approach \cite{DavidTomo} for the Wigner function of the {\it density operator}  is presented in Sec.~\ref{chapt:summary_reconstr}.
The generalized methodology for sampling spherical functions
representing the Wigner function of {\it propagators} is introduced in Sec.~\ref{theory}, which also
states the main technical results.
The experimental protocol and its implementation on an NMR spectrometer
are detailed in Secs.~\ref{NMR_implementation} and \ref{experiments}. The results of the NMR experiments
are summarized in Sec.~\ref{chapt:results}. 
We conclude in Secs.~\ref{sec:discussion} and \ref{sec:conclusion} by 
summarizing and discussing theoretical 
and experimental 
aspects and with an outlook on possible extensions of the presented approach.
Further technical details and illustrative examples are deferred to the Appendix.

\section{Visualization of operators using spherical functions}
\label{chapt:summary_drops}
\begin{figure}
\begin{center}
\includegraphics{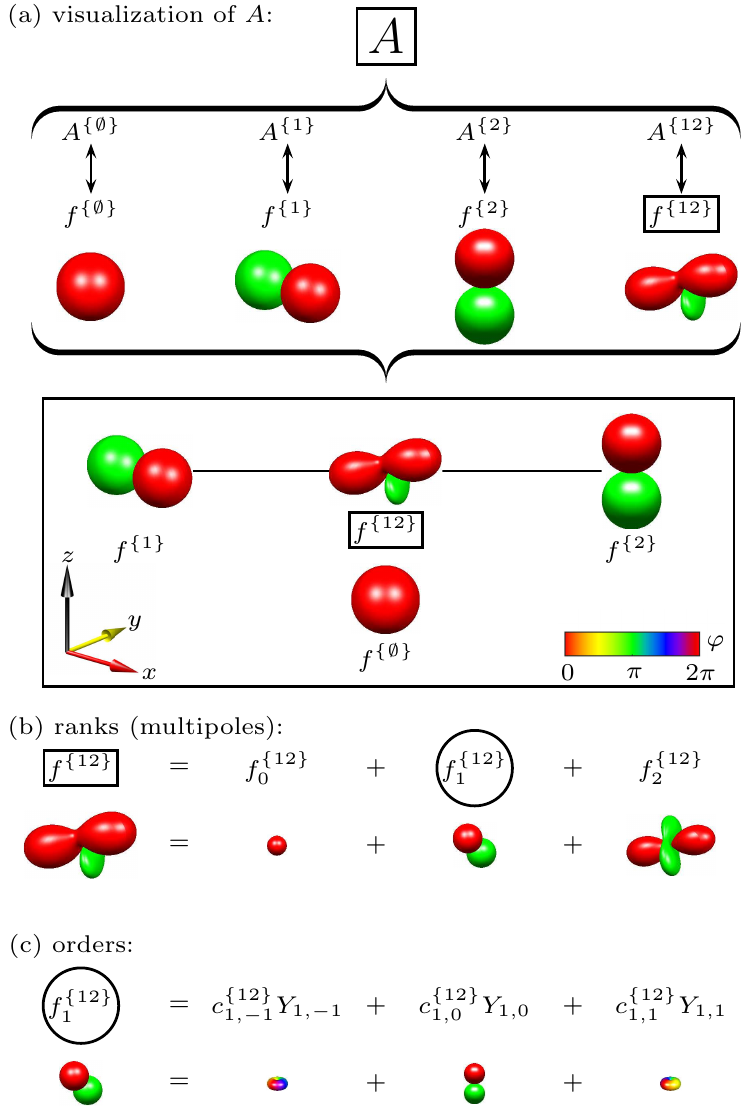}
\caption{(Color online) (a) The two-spin operator
$A=\tfrac{1}{2}{\bf 1} + I_{1x}+I_{2z}+I_{1x}I_{2x}+I_{1x}I_{2y} +
I_{1x}I_{2z}$
\cite{Note3,EBW87}
is represented using multiple spherical functions $f^{(\ell)}=f^{(\ell)}(\theta,\phi)$,
and individual components $A^{(\ell)}$ of $A$ mapped 
to  $f^{(\ell)}$ and graphically visualized.
(b)~$f^{\{12\}}$ (box)
decomposed into its
$2^j$-multipole contributions $f^{\{12\}}_j$ 
with $j \in \{0,1,2\}$. 
(c)~$f^{\{12\}}_1$ (circle) decomposed
into spherical harmonics of order $m\in\{-1,0,1\}$;  $Y_{1,-1}$ and $Y_{1,1}$ 
are rainbow colored \cite{Note4}.
\label{fig:drops}} 
\end{center}
\end{figure}

For a system consisting of a single spin, any spin operator can be mapped bijectively to a {\it single} (in general complex) spherical function using the Wigner representation \cite{Stratonovich,Garon15}.
This is achieved by expressing the operator as a linear combination of spherical tensor operators and mapping the spherical tensor operators to the corresponding 
spherical harmonics
\cite{Stratonovich}.
For a system of coupled spins, this approach to map an arbitrary operator to a  {\it single} spherical function is in general not bijective.
However, in this case a bijective Wigner representation can still be found if a spin operators $A$ is not mapped to a single spherical function but to  a discrete  {\it set} of
spherical functions,  called the DROPS representation \cite{Garon15}.
The individual spherical functions are called droplet functions or simply {\it droplets}  and are denoted 
$f^{(\ell)}=f^{(\ell)}(\theta,\phi)$ with $\ell \in L$, where $L$ is a set of labels $\ell$. The angles $\theta$ and $\phi$ 
are polar and azimuthal angles, respectively.
This mapping requires decomposing the operator $A$ into a sum of a corresponding discrete  {\it set} of droplet operators $A^{(\ell)}$:
\begin{equation}
\label{A}
A = \sum_{\ell \in L} A^{(\ell)}.
\end{equation}
As discussed in more detail  in \cite{Garon15}, many different partitions are possible but here we focus on the
so-called LISA basis, which characterizes each droplet operator
$A^{(\ell)}$ uniquely by its \underline{li}nearity (i.e., the number of involved spins) and the {\underline s}ubsystem (i.e., the identity of the involved spins).
For three or more spins, additional
 {\underline a}uxiliary criteria, such as permutation symmetry, etc., are needed to uniquely label the droplets \cite{Garon15}.
This is illustrated in Fig. \ref{fig:drops} (a) for the simple case of two coupled spins 1/2 denoted $I_1$ and $I_2$. In this case,
the set of droplet operators $A^{(\ell)}$ consists of four elements with labels 
$\{ 1 \}$, $\{ 2 \}$,  $\{ 12 \}$, and $\{\emptyset\}$.
The linear droplet operators 
$A^{\{ 1 \}}$ and $A^{\{ 2 \}}$ 
act only on spins $I_1$ and $I_2$, respectively. 
The bilinear droplet operator
$A^{\{{ 12 \}}}$
acts on both spins $I_1$ and $I_2$,
whereas the droplet operator $A^{\{\emptyset\}}$ is proportional to the identity operator {\bf 1} and acts neither on spin $I_1$ nor on spin $I_2$.

As shown in the center of Fig. \ref{fig:drops} (a), the operators 
$A^{\{ 1 \}}$, $A^{\{ 2 \}}$, $A^{\{{ 12 \}}}$, and $A^{\{\emptyset\}}$
can be mapped to the droplet functions $f^{\{ 1 \}}$, $f^{\{ 2 \}}$,
$f^{\{ 12 \}}$, and $f^{\{\emptyset\}}$, which can be graphically displayed as colored three-dimensional shapes.
In these three-dimensional polar plots of the droplets $f^{(\ell)}(\theta,\phi)$, the distance from the origin to a point on the surface represents the absolute value
$|f^{(\ell)}(\theta,\phi)|$ and the color represents the phase $\varphi=\text{arg}[f^{(\ell)}(\theta,\phi)]$ as defined by the color bar.
At the bottom of Fig. \ref{fig:drops} (a), the four droplets are arranged such that the positions of the two linear droplets $f^{\{ 1 \}}$ and $f^{\{ 2 \}}$
correspond to the positions of the two spins $I_1$ and $I_2$. The bilinear droplet $f^{\{ 12 \}}$ is positioned in the center of the line connecting 
$f^{\{ 1 \}}$ and $f^{\{ 2 \}}$ and the droplet $f^{\{\emptyset\}}$ is plotted separately below the other droplets.
This graphical representation of the droplets closely corresponds to physical intuition and makes it easy e.g. to follow the dynamics 
of spin operators and the involved spins.

The droplet operators $A^{(\ell)}$ can be further decomposed into multipole
components $A_{j}^{(\ell)}$ with different ranks $j$, where each 
$A^{(\ell)}$
only contains a finite number of possible ranks:
\begin{equation}\label{Aj}
A^{(\ell)}  = \sum_{j \in J(\ell)} A_{j}^{(\ell)}.
\end{equation}
The set $J(\ell)$ includes all occurring ranks $j$ of a given droplet $\ell$ \cite{Garon15}.
For example, for the bilinear droplet operator $A^{\{{ 12 \}}}$
possible values of the rank $j$ are 0, 1, and 2, for the linear droplet operators $A^{\{ 1 \}}$ and $A^{\{ 2 \}}$ the only possible rank is $j=1$ and for 
$A^{(\emptyset)}$ the only possible rank is $j=0$ \cite{Garon15}.

The decomposition of each droplet operator
$A^{( \ell )}$ is reflected by a decomposition of the corresponding droplet function $f^{(\ell)}$:
\begin{equation}\label{Ajb}
f^{(\ell)}  = \sum_{j \in J(\ell)}  f_{j}^{(\ell)}.
\end{equation}
This is illustrated in Fig. \ref{fig:drops} (b) for the decomposition of  $f^{\{ 12 \}}$
into the three rank-$j$ droplet functions $f_0^{\{ 12 \}}$, $f_1^{\{ 12 \}}$, and $f_2^{\{ 12 \}}$.

Finally, the rank-$j$ droplet operators $A^{(\ell)}_j$
can be decomposed into a linear combination of
components of irreducible spherical
tensor operators $T_{jm}^{(\ell)}$ \cite{Wigner31,Wigner59,Racah42,BL81,Silver76,CH98}
of order $m$ with $-j\leq m\leq j$ and (in general complex) coefficients $c_{jm}^{(\ell)}$.
It is at this level, that the DROPS representation exploits the 
well-known mathematical correspondence between 
irreducible tensor operators  and spherical harmonics $Y_{jm}=Y_{jm}(\theta, \phi)$ \cite{Silver76,CH98}.
Mapping the operators $T_{jm}^{(\ell)}$ to the corresponding functions $Y_{jm}$
results in the following mapping between
the rank $j$ component $A^{(\ell)}_j $ of the operator $A^{(\ell)}$ to the 
rank $j$ component of $f^{(\ell)}_j$ of the droplet function $f^{(\ell)}$:
\begin{equation}\label{Aj_fjd}
A^{(\ell)}_j  = \sum_{m=-j}^{j} c_{jm}^{(\ell)} T_{jm}^{(\ell)}
\ \  \longleftrightarrow
\ \ f^{(\ell)}_j  = \sum_{m=-j}^{j} c_{jm}^{(\ell)} Y_{jm},
\end{equation}
where the coefficients $c_{jm}^{(\ell)}$ in the left and the right sums are identical.
Figure. \ref{fig:drops} (c) illustrates the synthesis of the bilinear rank 1 droplet function $f_1^{\{ 12 \}}$
based on the linear combination of the spherical harmonics 
$Y_{1,-1}$, $Y_{1,0}$, and $Y_{1,1}$ with coefficients $c_{1,-1}^{\{ 12\}}$, $c_{1,0}^{\{ 12\}}$, and $c_{1,1}^{\{ 12\}}$.

Overall, this results in a bijective mapping between each droplet operator $A^{(\ell)}$ and the corresponding
spherical droplet functions $f^{(\ell)}$,
\begin{equation}\label{Ajc}
A^{(\ell)}  = \sum_{j \in J(\ell)} A_{j}^{(\ell)} \ \longleftrightarrow
\ \  f^{(\ell)}  = \sum_{j \in J(\ell)} f_{j}^{(\ell)}
\end{equation}
and in the mapping of any arbitrary operator $A$ to a corresponding set of droplet functions $f^{(\ell)}$,
\begin{equation}\label{mastermap}
A  = \sum_{\ell \in L}  A^{(\ell)} \   \longleftrightarrow
\ \  \bigcup_{\ell \in L}  f^{(\ell)}.
\end{equation}

\section{Summary of the scanning approach for density operators}
\label{chapt:summary_reconstr}
We summarize the experimental reconstruction approach of \cite{DavidTomo} to obtain a Wigner representation of a density operator.
Consider an arbitrary multi spin operator $A$, whose Wigner representation 
corresponds to a set of
spherical droplet functions  $f^{(\ell)}(\theta,\phi)=\sum_{j \in J(\ell)} 
f_{j}^{(\ell)}(\theta,\phi)$ as introduced in Sec. \ref{chapt:summary_drops}. 
The angles $\theta$ and $\phi$ indicate generic argument values of a spherical function
$g(\theta,\phi)$, whereas in the following the angles $\alpha$ and $\beta$ will be used to refer to specific argument values. 
In order to distinguish matrices of different size, in the following we use the notation $A^{[N]}$, where the 
superscript $[N]$ indicates the number of spins in the spin system. 
The label $\ell$ discriminates between the different spherical droplet functions. For each label $\ell$, the rank-$j$ component $f_{j}^{(\ell)}(\beta,\alpha)$ of $f^{(\ell)}  = \sum_{j \in J(\ell)} f_{j}^{(\ell)}$
can be determined 
for  polar  angles $0\leq \beta\leq \pi$ and azimuthal angles $0 \leq \alpha <
2\pi$ by
\begin{equation}
\label{tomoprime}
f_j^{(\ell)}(\beta, \alpha)   =   s_j\;
\langle {T}_{j, \alpha \beta}^{(\ell)[N]}| A^{[N]}\rangle
\end{equation}
with the scalar product
\begin{equation} \label{scalprod}
\langle {T}_{j, \alpha \beta}^{(\ell)[N]} | A^{[N]}\rangle=
{\rm tr}\{ ({T}_{j, \alpha \beta}^{(\ell)[N]})^\dagger A^{[N]} \},
\end{equation}
and where 
\begin{equation}
\label{defT}
{T}_{j, \alpha \beta}^{(\ell)[N]}= 
R^{[N]}_{\alpha\beta} (T_{j0}^{(\ell)})^{[N]} (R^{[N]}_{\alpha\beta})^\dagger 
\end{equation}
is the rotated version of an
axial tensor
operator $(T_{j0}^{(\ell)})^{[N]}$ of rank $j$ and order $0$ as introduced in Sec. \ref{chapt:summary_drops} (see also Result 2 in \cite{DavidTomo}). The rotation operator $R^{[N]}_{\alpha\beta}={\rm exp}\{-
{\rm{i}} \alpha F^{[N]}_z\}{\rm exp}\{-{\rm{i}} \beta F^{[N]}_y\}$
with the total spin operators $F^{[N]}_z=\sum_{k=1}^N I^{[N]}_{kz}$ and $F^{[N]}_y=\sum_{k=1}^N I^{[N]}_{ky}$
corresponds to rotation around the $y$ axis by $\beta$ followed by rotation around the $z
$ axis by $\alpha$. 
The operators $I^{[N]}_{kb}$ with ${b} \in \{x,y,z \}$ are spin operators acting only
on the $k$-th spin $I_k$ \cite{EBW87}.
The prefactors $s_j$ are given by $s_j=\sqrt{(2j{+}1)/(4\pi)}$.

If the density matrix $\rho^{[N]}$ of a spin system can be prepared to be identical to the 
operator $A^{[N]}$,
for all droplets $\ell$ the rank-$j$ droplet components $f_{j}^{(\ell)}$ for $j \in J(\ell)$ become experimentally accessible 
by measuring expectation 
values \cite{DavidTomo}

\begin{equation} 
\label{expvals}
f_j^{(\ell)}(\beta, \alpha)   =   s_j\;
\langle  {T}_{j, \alpha \beta}^{(\ell)[N]} \rangle_{\rho^{[N]}},
\end{equation}
with the expectation value given by
\begin{equation} 
\label{expvals1}
\langle  {T}_{j, \alpha \beta}^{(\ell)[N]} \rangle_{\rho^{[N]}} = {\rm tr}\{ {T}_{j, \alpha \beta}^{(\ell)[N]}\rho^{[N]}\}.
\end{equation}

This is due to the fact that
the expectation value in Eq. \eqref{expvals1} is identical to the scalar products $ \langle {T}_{j, \alpha \beta}^{(\ell)[N]} | A^{[N]}\rangle$ 
of Eq. (\ref{scalprod}),
since the rotated axial tensor operators are Hermitian and hence 
$({T}_{j, \alpha \beta}^{(\ell)[N]})^\dagger={T}_{j, \alpha \beta}^{(\ell)[N]}$. 

Equation \eqref{expvals} states that the value of the rank-$j$
droplet components $f_j^{(\ell)}(\beta, \alpha)$ 
for a density matrix $\rho^{[N]}$
can be calculated from expectation values of rotated axial tensor operators
$\langle  {T}_{j, \alpha \beta}^{(\ell)[N]} \rangle_{\rho^{[N]}}$
and further implies that one can then retrace the shapes of the spherical function
$f^{(\ell)}=\sum_{j \in J(\ell)} 
f_{j}^{(\ell)}$
representing $\rho^{[N]}$
if one experimentally measures $f^{(\ell)}(\beta,\alpha)$
as a function of $\alpha$ and $\beta$.

\section{Theory of the Wigner process tomography}
\label{theory}
We are interested in experimentally scanning the shape of the Wigner representation of a 
propagator $U^{[N]}$, i.e., of the unitary
time-evolution operator created by a given pulse sequence. This is of interest to 
characterize an unknown propagator or to test how well a propagator $U^{[N]}$ created by an experimentally  implemented pulse sequence 
approaches a desired propagator $U^{[N]}_{target}$. The targeted evolution operator $U^{[N]}_{target}$ 
could be a spectroscopically relevant propagator, which is, e.g., designed to create multiple-quantum coherence from thermal equilibrium spin polarization \cite{Köcher2016},
or a target 
propagator corresponding to a quantum gate, or an entire quantum algorithm \cite{NC00}.

We consider a system consisting of $N$ spins  $I_k$ for $1\leq k \leq N$.
For simplicity, but without loss of generality, here we assume that all spins are spin 1/2-particles (qubits). In this case, the size of the Hilbert space is $2^{N}$ and operators 
$A^{[N]}$ in this Hilbert space (such as the density operator $\rho^{[N]}$ or propagators 
$U^{[N]}$) are represented by $2^{N} \times 2^{N} $ matrices.
If the operator of interest is the {\it density operator}, $\rho^{[N]}$, of the spin system, 
according to Eq. \eqref{expvals}
the spherical functions $f^{(\ell)}_j (\beta,\alpha)$ representing droplet components
can be determined experimentally by
measuring the expectation values of the operators ${T}_{j, \alpha \beta}^{(\ell)[N]}$. 
Hence, it is possible to scan the DROPS representation of an arbitrary operator $A^{[N]}$ if 
it can be experimentally mapped onto the density operator.
However, as the density operator $\rho^{[N]}$ is a {\it Hermitian} matrix, whereas  the 
propagator $U^{[N]}$ is represented by a {\it unitary} matrix, it is in general not possible 
to map the $2^{N} \times 2^{N} $ matrix representations of an arbitrary propagator $U^{[N]}$  
on a $2^{N} \times 2^{N} $  matrix representing a density operator $\rho^{[N]}$ and Eqs.~\eqref{expvals1} and \eqref{expvals} are not directly applicable for the tomography of the Wigner functions of unitary operators. However, as shown in 
the following, this roadblock to experimentally scanning the DROPS representation of a propagator 
$U^{[N]}$ can be lifted if the spin system is augmented by adding an {\it ancilla spin}  $I_0$.

\subsection{Inscribing $U^{[N]}$ on the density operator $\rho^{[N+1]}$ of a system augmented by an ancilla spin}
\label{sec:inscribe}
In the augmented system consisting of $N+1$ spins, all operators $A^{[N+1]}$ are represented 
by $2^{N+1}\times 2^{N+1} $ matrices. 
To simplify the notation and without loss of generality, in the following we characterize the state of the augmented system consisting of 
$N+1$ spins 1/2 by the traceless part of the full density operator \cite{EBW87, Fahmy2008, DavidTomo}, 
 denoted the {\it deviation density matrix} \cite{NC00}.
The deviation density matrix   $\rho^{[N+1]}$  (with trace zero) is obtained
from the full density operator (with trace one) by subtracting the matrix $(1/2^{N+1})\  {\bf 1}^{[N+1]}$, which is proportional to the $2^{N+1} \times 2^{N+1}$ identity matrix and each diagonal element is given by $(1/2^{N+1})$. The term proportional to the identity operator 
does not evolve under unitary transformations. Furthermore, in magnetic resonance experiments all detectable operators
are traceless. Hence the term proportional to the identity operator does not give rise to detectable signal and can be ignored \cite{NC00}.

It is possible to inscribe a propagator $U^{[N]}$ and 
its adjoint $(U^{[N]})^\dagger$ of the original system consisting of $N$ spins into the off-diagonal sub blocks of the density operator $\rho^{[N+1]}$ of the augmented system  
in the following way \cite{Myers2001, Fahmy2008, Marx2010}.

First, it is possible to design a pulse sequence which creates the
 {\it controlled} propagator $cU^{[N+1]}$ \cite{Barenco1995, Fahmy2008, Marx2010}, 
 which for spins $I_1$, ..., $I_N$ has no effect if the ancilla spin $I_0$ is in 
the {\it up} state $|\hspace{-1mm} \uparrow\rangle$ but creates 
the propagator $U^{[N]}$ if the ancilla spin $I_0$ is in the {\it down} state $|\hspace{-1mm}\downarrow\rangle$. Hence, the $2^{N+1} \times 2^{N+1} $ matrix 
representing
 the propagator $cU^{[N+1]}$ is block-diagonal. The first block corresponds to the $2^{N} 
\times 2^{N} $-dimensional identity matrix {\bf 1}$^{[N]}$ and the second block is the $2^{N} 
\times 2^{N} $-dimensional propagator $U^{[N]}$:
 \begin{equation}\label{rcu0}
 cU^{[N+1]} = 
\begin{pmatrix} {\bf 1}^{[N]} & {\bf 0}^{[N]} \\ {\bf 0}^{[N]} & U^{[N]}\end{pmatrix},
 \end{equation}
 where ${\bf 0}^{[N]}$ is the $2^{N} \times 2^{N} $-dimensional zero matrix.
 In the field of quantum information processing, the spin $I_0$ is called the {\it control 
qubit} and the spins $I_1,\dots, I_N$ on which the unitary operator $U^{[N]}$ act are 
called the {\it target qubits} \cite{NC00}. 

Second, by putting the ancilla spin $I_0$ into a superposition state $(|\hspace{-1mm} \uparrow \rangle + |\hspace{-1mm} \downarrow\rangle )/\sqrt{2}$ 
and putting the remaining spins into a fully mixed state,
the  system can be prepared such that the deviation density 
operator is proportional to
\begin{equation}\label{rh}
\rho^{[N+1]}_0=2 I^{[N+1]}_{0x}=2 I_x \otimes {\bf 1}^{[N]}.
\end{equation}
Here the spin operators $I_{b}$ are defined as $I_{b} := 1/2 \ \sigma_{b}$ \cite{EBW87,Note3} for ${b} \in \{x,y,z \}$, where $\sigma_{b}$ is a Pauli spin 
operator.
In NMR implementations, this is achieved by 
exciting the ancilla spin $I_0$, i.e., rotating its thermal equilibrium Bloch vector to the $x$ direction
and saturating the remaining spins e.g. by a combination of pulses and pulsed $B_0$ gradients (see Sec. \ref{exp_protocol}).

With these ingredients, we can imprint the propagator $U^{[N]}$ on the density operator by 
applying the controlled propagator $cU^{[N+1]}$ to 
$\rho_0^{[N+1]}$. This prepares the deviation density matrix \cite{Myers2001, Fahmy2008, Marx2010}
\begin{equation}\label{rhoU}
\rho^{[N+1]}_U=cU^{[N+1]}\  \rho^{[N+1]}_0 \ (cU^{[N+1]})^{\dag} = \begin{pmatrix} {\bf 0}
^{[N]} & (U^{[N]})^{\dag} \\ U^{[N]} & {\bf 0}^{[N]} \end{pmatrix},
\end{equation}
which can be expressed in the form 
\begin{equation}\label{rhoU1}
\rho_U^{[N+1]}= I^- \otimes U^{[N]} + I^+ \otimes (U^{[N]})^\dagger
\end{equation}
using the raising and lowering operators $I^+=I_x+ {i}I_y$ and $I^-=I_x- {i}I_y$ \cite{EBW87}.
In Appendix \ref{append_example_cU}, explicit matrices of the operators $U^{[N]}$, 
$cU^{[N+1]}$, $\rho_0^{[N]}$ and $\rho_U^{[N+1]}$ are provided for a simple example, where 
the system of interest consists only of a single spin $I_1$, augmented by an ancilla 
spin $I_0$. 
Note that the two non-zero  off-diagonal sub blocks of $\rho_U^{[N+1]}$ are unitary (and in 
general non-Hermitian), but the overall deviation density operator  is Hermitian, i.e. $\rho^{[N+1]}
_U=(\rho^{[N+1]}_U)^\dagger$.

\subsection{Scanning the DROPS Wigner representation of $U$ based on the density operator of the augmented spin system}

\begin{figure}
\begin{center}
\includegraphics{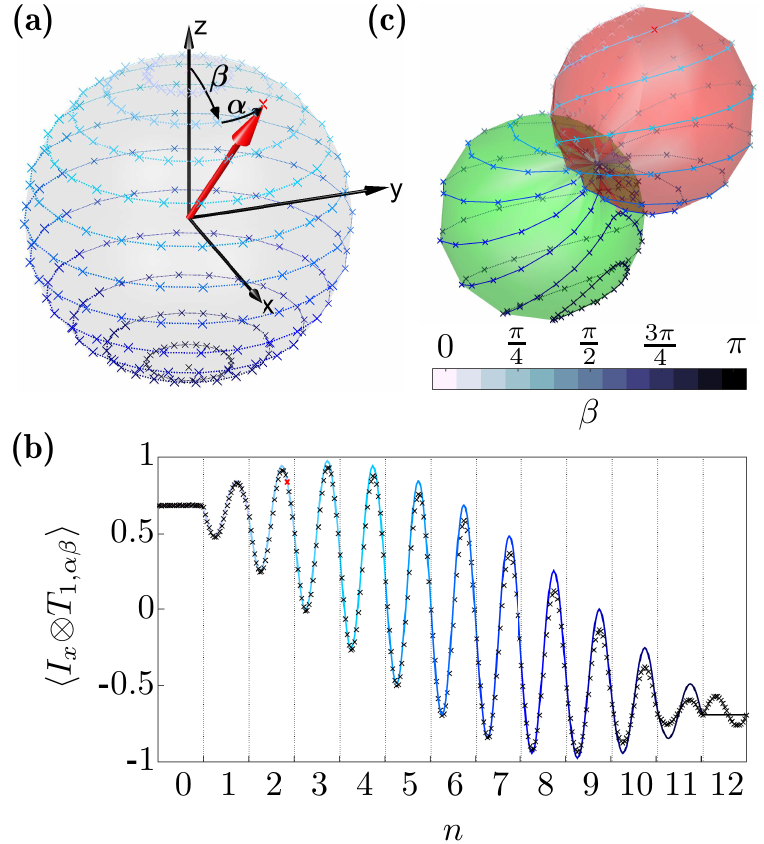}
\caption{ (Color online) For a system of interest consisting of a single ($N=1$) spin 1/2, the figure illustrates the tomographic
reconstruction of a spherical function  $f^{(\ell)}(\beta,\alpha)$ representing the Hadamard gate based on Result~\ref{eq:prop_reconstr}. (a) Samples (crosses) with different polar
angles $\beta$ in the range $[0,\pi]$  and
phases $\alpha$ in the range $[0,2\pi]$ acquired by expectation values $\langle  I_a  {\otimes} 
{T}_{j, \alpha \beta}^{(\ell)[N]} \rangle_{\rho^{[N+1]}_U}$ with $a \in \{x,y\}$ and $j \in \{0,1\}$.
The colors (shades of gray) of the circles of latitude are defined as a function of the polar angle $\beta$ by the given color bar (grayscale).
(b) Predicted expectation values (line) and experimentally measured  expectation values $\langle  I_x  {\otimes}  
{T}_{1, \alpha \beta} \rangle :=\langle  I_x  {\otimes} 
{T}_{1, \alpha \beta}^{(\ell)[N]} \rangle_{\rho^{[N+1]}_U}$ (crosses) 
for a simple sampling scheme with an equidistant discrete set of polar angles $\beta \in \{0,\tfrac{\pi}{12},\tfrac{2\pi}{12},\dots,\pi\}$ and azimuthal angles $\alpha \in \{0,\tfrac{\pi}{12},\tfrac{2\pi}{12},\dots,2\pi\}$. 
For each discrete value  $\beta_n =  n \pi/12$   with $n \in \{0,1,2,\dots,12\}$, 
the azimuthal angle $\alpha$ is incremented in steps of $\pi/12$ from 0 to $2 \pi$.
This scheme results in $13\cdot25=325$ measurement points acquired in 13 cycles to fully scan the sphere.
Note that the remaining expectation values $\langle  I_x  {\otimes} 
{T}_{0, \alpha \beta}^{(\ell)[N]} \rangle_{\rho^{[N+1]}_U}$, $\langle  I_y  {\otimes} 
{T}_{0, \alpha \beta}^{(\ell)[N]} \rangle_{\rho^{[N+1]}_U}$, and $\langle  I_y  {\otimes} 
{T}_{1, \alpha \beta}^{(\ell)[N]} \rangle_{\rho^{[N+1]}_U}$ are zero and not shown here.
(c) Smooth surface interpolated
from individual samples with distance from the origin given by
$f^{(\ell)}(\beta,\alpha)$, the phase of which determines the color (grayscale) of the surface
(see Fig. \ref{fig:1qubitgates} ).}
\label{fig:reconstr} 
\end{center}
\end{figure}

As shown in Sec.~\ref{chapt:summary_reconstr}, and in more detail in \cite{DavidTomo}, the key to the scanning approach of the DROPS 
Wigner representation of operators is the
experimental determination of scalar products between rotated axial tensor operators ${T}_{j, \alpha \beta}^{(\ell)[N]}$ and the 
operator of interest $A^{[N]}$
\begin{equation}
\langle {T}_{j, \alpha \beta}^{(\ell)[N]} | A^{[N]}\rangle=\langle  {T}_{j, \alpha \beta}^{(\ell)[N]}
\rangle_{\rho^{[N]}},
\end{equation}
if $\rho^{[N]}=A^{[N]}$.
As shown in Appendix \ref{proof}, for $A^{[N]}=U^{[N]}$ and $\rho_U^{[N+1]}$, the corresponding 
scalar products
can be obtained by the following complex linear combination of expectation values of the 
Hermitian operators 
$I_x  \otimes {T}_{j, \alpha \beta}^{(\ell)[N]}$ and $I_y  \otimes {T}_{j, \alpha \beta}^{(\ell)[N]}$:
\begin{equation}
\label{eq:scalU}
\langle{T}_{j, \alpha \beta}^{(\ell)[N]} | U^{[N]}\rangle=  
\langle  I_x  \otimes {T}_{j, \alpha \beta}^{(\ell)[N]}\rangle_{\rho^{[N+1]}_U} 
+ {i} \ \langle  I_y  \otimes {T}_{j, \alpha \beta}^{(\ell)[N]} \rangle_{\rho^{[N+1]}_U}. 
\end{equation}
In Appendixes \ref{append_example_cU} and \ref{append_example_meas}, this is also illustrated 
for a simple example. Thus, an analog formula to Eq.~\eqref{expvals} for the reconstruction of Wigner functions 
representing propagators is given by the following result.
\begin{result}
\label{eq:prop_reconstr}
Consider a propagator $U^{[N]}$ which is represented by a set of spherical functions 
$f^{(\ell)}(\theta,\phi) = \sum_{j \in J(\ell)} f_j^{(\ell)}(\theta,\phi)$. The rank-$j$ components 
$f_j^{(\ell)}(\beta,\alpha)$ can be experimentally measured in the augmented system for 
arbitrary angles $\beta$ in the range $[0,\pi]$ and $\alpha$ in the range $[0,2\pi]$ via the complex linear combination of expectation 
values 
\begin{equation}
\label{eq:prop_reconstr_eq}
f_j^{(\ell)} (\beta, \alpha)= s_j \left( \langle  I_x  {\otimes} {T}_{j, \alpha \beta}^{(\ell)[N]}
\rangle_{\rho^{[N+1]}_U} 
{+} {\rm{i}}  \langle  I_y  {\otimes} {T}_{j, \alpha \beta}^{(\ell)[N]} \rangle_{\rho^{[N+1]}_U} 
\right)
\end{equation}
with $s_j := \sqrt{(2j+1)/(4\pi)}$ and, to simplify notation, where ${T}_{j, \alpha \beta}^{(\ell)[N]}= R^{[N]}_{\alpha\beta} (T_{j0}^{(\ell)})^{[N]} 
(R^{[N]}_{\alpha\beta})^\dagger$ is a rotated axial tensor
operator $(T_{j0}^{(\ell)})^{[N]}$. The rotation operator $R^{[N]}_{\alpha\beta}={\rm exp}\{-
{\rm{i}} \alpha \sum_{k=1}^N I^{[N]}_{kz}\}{\rm exp}\{-{\rm{i}} \beta \sum_{k=1}^N I^{[N]}_{ky}\}$
corresponds to a rotation around the $y$ axis by $\beta$ followed by a rotation around the $z
$ axis by $\alpha$ as shown in Fig.~\ref{fig:reconstr}.
\end{result}

A general procedure to reconstruct the spherical functions $f^{(\ell)}(\beta,\alpha)$ representing a 
propagator $U^{[N]}$ for the system of interest $I_1,\dots, I_N$ is first to prepare the 
augmented system $I_0,\dots, I_N$ in the state $\rho_0^{[N+1]}$ and apply the controlled 
unitary $cU^{[N+1]}$ to $\rho_0^{[N+1]}$ to generate $\rho^{[N+1]}_U$ 
[see Eqs.~\eqref{rcu0}-\eqref{rhoU}]. Then, for all droplets $\ell$ according to 
Result~\ref{eq:prop_reconstr}, the shape of $f^{(\ell)}(\beta, \alpha)$ can be scanned by measuring the expectation values $\langle  I_x  {\otimes} 
{T}_{j, \alpha \beta}^{(\ell)[N]} \rangle_{\rho^{[N+1]}_U}$ and $\langle  I_y  {\otimes} {T}_{j, \alpha \beta}^{(\ell)[N]} \rangle_{\rho^{[N+1]}_U}$ for all $j \in J(\ell) $ 
as a function of the angles $\alpha$ and $\beta$ as exemplified in Fig~\ref{fig:reconstr}.
Different sampling schemes for $\alpha$ and $\beta$ are discussed in Sec. \ref{experiments} B (see also Figs. \ref{fig:reconstr} and \ref{fig:exp}) and  in Sec. \ref{sec:discussion}.

\section{NMR implementation}
\label{NMR_implementation}
For simplicity, but without loss of generality, let us assume that the ancilla spin $I_0$ is 
coupled to all spins of the system of interest ($I_1$,...,$I_N$). We will now outline the 
experimental implementation of Result~\ref{eq:prop_reconstr} using methods of nuclear 
magnetic resonance. 
We start in Sec.~\ref{impl_NMR} by presenting an NMR-adapted version of Result~
\ref{eq:prop_reconstr}. An example for the case $N=1$ will be given in Sec.~\ref{caseN1}. The 
realization of controlled propagators $cU^{[N+1]}$ and the implementation of rotation operations are 
outlined in Secs.~ \ref{impl_cU} and \ref{impl_rot}.

\subsection{NMR-based reconstruction}
\label{impl_NMR}
The expectation values $\langle  I_x  {\otimes} {T}_{j, \alpha \beta}^{(\ell)[N]}\rangle_{\rho^{[N+1]}_U}$ and $\langle  I_y  {\otimes} {T}_{j, \alpha \beta}^{(\ell)[N]} \rangle_{\rho^{[N+1]}_U}$ of Result~\ref{eq:prop_reconstr} are not directly obtainable with methods of nuclear magnetic resonance. An
NMR-adapted version of Result~\ref{eq:prop_reconstr} will be presented in the following.
This NMR-based experimental reconstruction scheme of propagators is analogous to the scheme
described in \cite{DavidTomo} for the reconstruction of density operators.
First, instead of rotating the axial tensors $(T_{j0}^{(\ell)})^{[N]}$, the density matrix $
\rho_U^{[N+1]}$ is rotated inversely, resulting in
\begin{align}
\label{eq:rot_exp_1}
\langle  I_x  {\otimes}  {T}_{j, \alpha \beta}^{(\ell)[N]}
 \rangle_{\rho^{[N+1]}_U} &= \langle  I_x  
{\otimes} (T_{j0}^{(\ell)})^{[N]}  \rangle_{\tilde{\rho}^{[N+1]}_U} \\
\label{eq:rot_exp_2}
\langle  I_y  {\otimes}  {T}_{j, \alpha \beta}^{(\ell)[N]}
 \rangle_{\rho^{[N+1]}_U} &= \langle  I_y  
{\otimes} (T_{j0}^{(\ell)})^{[N]}  \rangle_{\tilde{\rho}^{[N+1]}_U}
\end{align}
with 
\begin{align}
\label{eq:rot_rho}
\tilde{\rho}^{[N+1]}_U = (\tilde R^{[N+1]}_{\alpha\beta})^{-1} \rho^{[N+1]}_U \tilde R^{[N+1]}_{\alpha\beta}
\end{align}
and with $\tilde R^{[N+1]}_{\alpha\beta} = {\bf 1}^{[1]} \otimes R^{[N]}_{\alpha\beta}$, 
as discussed in more detail in Appendix \ref{app:proof_inv_rot}. \\
Secondly, the (Hermitian) rank-$j$ tensor components $(T_{j0}^{(\ell)})^{[N]}$
can be decomposed into (Hermitian) Cartesian product operators \cite{EBW87}
$(C_{j,n}^{(\ell)})^{[N]}$
via
\begin{equation}
\label{eq:trafo_T}
(T_{j0}^{(\ell)})^{[N]}  = \sum_n r_{j,n}^{(\ell) [N]} (C_{j,n}^{(\ell)})^{[N]}
\end{equation}
with real expansion coefficients $r_{j,n}^{(\ell) [N]} \in \mathbb{R}$.
The decompositions of relevant axial tensor components for up to three qubits were given explicitly in 
\cite{DavidTomo}. 
Hence, the expectation values of Eqs. \eqref{eq:rot_exp_1} and \eqref{eq:rot_exp_2} can be obtained if we can measure the expectation values of the 
Cartesian product operators $I_a  {\otimes}  (C_{j,n}^{(\ell)})^{[N]}$ with $a \in \{x,y\}$.
As in NMR experiments 
the signatures of Cartesian product operators 
can be measured directly only if
they contain exactly  {\it one}
{\it transverse} Cartesian spin operator $I_{ka}$;
 the expectation values $\langle I_a {\otimes}  (C_{j,n}^{(\ell)})^{[N]}\rangle_{\tilde{\rho}^{[N+1]}_U}$
are only measurable directly if 
the terms $(C_{j,n}^{(\ell)})^{[N]}$
do not contain any transverse Cartesian spin operator.
If this is not the case,
the Cartesian product operators $ I_a  {\otimes}  (C_{j,n}^{(\ell)})^{[N]}$ are transformed into NMR-measurable operators \cite{EBW87} $ (M^{(\ell)}_{j,n})^{[N+1]}_a \in \{I_{0a}, 2I_{0a}I_{1z}, 4I_{0a}I_{1z}I_{2z}, \dots \}$ 
by applying unitary operations
\begin{align}
\label{trafo1}
 (M^{(\ell)}_{j,n})^{[N+1]}_a &= ( \tilde V_{j,n}^{(\ell)})^{[N+1]} \left (I_a \otimes (C_{j,n}^{(\ell)})^{[N]} \right ) [( \tilde V_{j,n}^{(\ell)})^{[N+1]}]^\dagger
\end{align}
with 
$(\tilde V_{j,n}^{(\ell)})^{[N+1]} = {\bf 1}^{[1]} \otimes (V_{j,n}^{(\ell)})^{[N]}$ 
and with the actions of this operation on an operator $A^{[N+1]}$ given by 
$({\tilde V}_{j,n}^{(\ell)})^{[N+1]} A^{[N+1]} [(\tilde V_{j,n}^{(\ell)})^{[N+1]}]^\dagger$ 
[see Table IV in \cite{DavidTomo} for some realizations of $( V_{j,n}^{(\ell)})^{[N]}$].
The unitary operators $( V_{j,n}^{(\ell)})^{[N]}$
can be experimentally implemented using radio-frequency pulses and evolution periods under 
couplings.
The experimental  signatures of the Cartesian product operators $(M^{(\ell)}_{j,n})^{[N+1]}_a$ are obtained by detecting the ancilla spin $I_0$.

This approach to measure spherical
functions corresponding to the DROPS representation of propagators recasts Result~\ref{eq:prop_reconstr} into the following NMR-adapted 
version:
\begin{result}
\label{eq:prop_reconstr_NMR}
Consider a propagator $U^{[N]}$ which is represented by a set of spherical functions 
$f^{(\ell)}(\theta,\phi) = \sum_{j \in J(\ell)} f_j^{(\ell)}(\theta,\phi)$. The rank-$j$ components 
$f_j^{(\ell)}(\beta,\alpha)$ can be experimentally measured with NMR methods in the augmented system for 
arbitrary angles $\beta$ in the range $[0,\pi]$ and $\alpha$ in the range $[0,2\pi]$ via the complex linear combination of expectation 
values 
\begin{align}
&f_j^{(\ell)} (\beta, \alpha) \\
&= \hspace{-1mm} s_j \sum_n r_{j,n}^{(\ell) [N]}\hspace{-1mm} \left(\hspace{-1mm} \langle  (M^{(\ell)}_{j,n})^{[N+1]}_x 
\rangle_{\tilde{\tilde{\rho}}^{[N+1]}_U} 
{+} {\rm{i}} \langle  (M^{(\ell)}_{j,n})^{[N+1]}_y \rangle_{\tilde{\tilde{\rho}}^{[N+1]}_U} \hspace{-1mm}\right)\nonumber
\end{align}
with $s_j := \sqrt{(2j+1)/(4\pi)}$ and Cartesian product operators $(M^{(\ell)}_{j,n})^{[N+1]}_a $ with $a \in \{x,y \}$ given in Eq.~\eqref{trafo1} in which only the ancilla qubit has a transversal component, where 
\begin{align}
\tilde{\tilde{\rho}}^{[N+1]}_U &= (V_{j,n}^{(\ell)})^{[N+1]} \tilde{\rho}^{[N+1]}_U [(V_{j,n}^{(\ell)})^{[N+1]}]^\dagger, \\
\tilde{\rho}^{[N+1]}_U &= (R^{[N+1]}_{\alpha\beta})^{-1} \rho^{[N+1]}_U R^{[N+1]}_{\alpha\beta},
\end{align}
and
\begin{align}
R^{[N+1]}_{\alpha\beta} &= {\bf 1}^{[1]} \otimes R^{[N]}_{\alpha\beta}, \\
R^{[N]}_{\alpha\beta}&={\rm exp}\{-{\rm{i}} \alpha \sum_{k=1}^N I^{[N]}_{kz}\}{\rm exp}\{-{\rm{i}} \beta 
\sum_{k=1}^N I^{[N]}_{ky}\}.
\end{align}
\end{result}
The rank-$j$ components $f_j^{(\ell)}(\beta,\alpha)$ of the spherical functions $f^{(\ell)}
(\beta,\alpha)$ representing the propagator $U^{[N]}$ can be sampled in NMR experiments by 
preparing the state $\tilde{\tilde{\rho}}^{[N+1]}_U$ in the augmented system and measuring a 
set of expectation values of suitable operators $(M^{(\ell)}_{j,n})^{[N+1]}_x$ and $(M^{(\ell)}_{j,n})^{[N+1]}_y$ as a function of the angles $\alpha$ and $\beta$. 

\subsection{The case $N=1$}
\label{caseN1}
If the system of interest is just a single spin $I_1$, i.e., $N=1$ and $\ell \in \{\emptyset,1\}$, we have rank $j=0$ for $\ell=\{\emptyset\}$: $j=0$ and rank $j=1$ for $\ell=\{1\}$. In both cases, only a single ($n=1$) Cartesian product operator is necessary to express the axial tensor operator components $(T_{00}^{(\emptyset)})^{[1]}= \frac{1}{\sqrt{2}} {\bf 1}
^{[1]} $ and $(T_{10}^{(1)})^{[1]}= \sqrt{2} I_{1z}^{[1]}$; see Eq.~\eqref{eq:trafo_T}.
Here, the right-hand sides of Eqs.~
\eqref{eq:rot_exp_1} and \eqref{eq:rot_exp_2} reduce to
\begin{align}
\langle I_{x}\otimes (T_{00}^{(\emptyset)})^{[1]} \rangle_{\tilde{\rho}^{[N+1]}_U} &= \langle  \frac{1}
{\sqrt{2}} I_{0x}^{[2]} \rangle_{\tilde{\rho}^{[N+1]}_U} \\
\langle I_{y}\otimes (T_{00}^{(\emptyset)})^{[1]} \rangle_{\tilde{\rho}^{[N+1]}_U} &= \langle \frac{1}
{\sqrt{2}} I_{0y}^{[2]} \rangle_{\tilde{\rho}^{[N+1]}_U} 
\end{align}
for $\ell=\{\emptyset\}$, and
\begin{align}
\langle I_{x}\otimes (T_{10}^{(1)})^{[1]} \rangle_{\tilde{\rho}^{[N+1]}_U} &= \langle \frac{1}
{\sqrt{2}}(2I_{0x}
I_{1z})^{[2]} \rangle_{\tilde{\rho}^{[N+1]}_U}\\
\langle I_{y}\otimes (T_{10}^{(1)})^{[1]} \rangle_{\tilde{\rho}^{[N+1]}_U} &= \langle \frac{1}
{\sqrt{2}}(2I_{0y}
I_{1z})^{[2]} \rangle_{\tilde{\rho}^{[N+1]}_U}.
\end{align}
for $\ell=\{1\}$. Note that $I_{0 a}^{[2]} := I_{a}\otimes {\bf 1}^{[1]}$ and $(2I_{0 a}I_{1z})^{[2]} := 
2I_{a} \otimes I_{z}$ for $a \in \{x,y\}$.
Since the signatures of the four Cartesian product operators $I_{0x}^{[2]}$, $I_{0y}^{[2]}$, 
$(2I_{0x}I_{1z})^{[2]}$, and $(2I_{0y}I_{1z})^{[2]}$ are already directly observable in NMR 
experiments, further transformations $(V_{j,1}^{(\ell)})^{[2]}$ [see Eq.~\eqref{trafo1}] 
are not necessary, i.e., $(V_{j,1}^{(\ell)})^{[2]} = {\bf 1}^{[2]}$ for all $\ell$ and $j$ and thus $\tilde{\tilde{\rho}}^{[2]}_U = \tilde{\rho}^{[2]}
_U$. Hence, based on Result~\ref{eq:prop_reconstr_NMR},
we only need to measure the two expectation values for $\ell=\{\emptyset\}$, 
\begin{align}
&\langle (M^{(\emptyset)}_{0,1})^{[2]}_x \rangle_{\tilde{\tilde{\rho}}^{[2]}_U} = \frac{1}
{\sqrt{2}} \langle I_{0x}^{[2]} 
\rangle_{\tilde{\rho}^{[2]}_U}\\
&\langle  (M^{(\emptyset)}_{0,1})^{[2]}_y \rangle_{\tilde{\tilde{\rho}}^{[2]}_U} =\frac{1}
{\sqrt{2}}\langle I_{0y}^{[2]} 
\rangle_{\tilde{\rho}^{[2]}_U}
\end{align}
and the two expectation values for $\ell=\{1\}$ 
\begin{align}
&\langle  (M^{(1)}_{1,1})^{[2]}_x \rangle_{\tilde{\tilde{\rho}}^{[2]}_U} =\frac{1}
{\sqrt{2}}\langle (2I_{0x}
I_{1z})^{[2]} \rangle_{\tilde{\rho}^{[2]}_U}\\ 
&\langle  (M^{(1)}_{1,1})^{[2]}_y \rangle_{\tilde{\tilde{\rho}}^{[2]}_U} =\frac{1}
{\sqrt{2}}\langle (2I_{0y}
I_{1z})^{[2]} \rangle_{\tilde{\rho}^{[2]}_U}
\end{align}
as a function of the angles $\alpha$ and $\beta$ to reconstruct the spherical functions $f^{(\ell)} (\beta,\alpha)$ for each $\ell \in \{\emptyset,1\}$.
This is achieved by measuring the spectrum of the ancilla qubit
$I_0$ and fitting reference spectra of the operators
$I_{0x}^{[2]}$, $I_{0y}^{[2]}$, $(2I_{0x}
I_{1z})^{[2]}$ and $(2I_{0y}
I_{1z})^{[2]}$ \cite{DavidTomo}.

\subsection{Implementation of controlled unitary transformation}
\label{impl_cU}
It is always possible to implement a controlled unitary propagator using unitary 
transformations in NMR \cite{Barenco1995}. These are 
realized by radio-frequency pulses and evolutions under couplings. The explicit matrix 
representations of $U^{[N]}$ and the corresponding pulse sequences for all considered propagators $cU^{[N+1]}$ are 
summarized in Table~\ref{tab:c_prob}. 

\subsection{Implementation of rotations}
\label{impl_rot}

The (inverse) rotation $(R_{\alpha \beta}^{[N+1]})^{-1}$ transforms the density operator after 
the controlled gate $\rho_{U}^{[N+1]}$ into $\tilde{\rho}_{U}^{[N+1]}$ in order to probe 
the corresponding spherical functions representing the propagator $U^{[N]}$ 
for the polar angles $\beta$ and azimuthal angles $\alpha$. This operation 
is implemented by radio-frequency pulses $[\beta]_{\alpha-\pi/2}$ with  
flip angle $\beta$ and
phase $(\alpha-\pi/2)$, which are simultaneously applied to all spins in the system of 
interest ($I_1$, ... $I_N$) but not to the ancilla spin $I_0$.

\section{NMR-experiments for $N=1$}
\label{experiments}
After summarizing the theory and the basis of NMR implementations for the Wigner process 
tomography of propagators in the
 previous sections \ref{theory} and \ref{NMR_implementation}, we will now outline the 
 experimental procedure which is directly based on Result~\ref{eq:prop_reconstr_NMR} using 
 methods of nuclear
 magnetic resonance for a system with one system qubit and one ancilla qubit, i.e., 
 $N=1$ (see Sec.~\ref{caseN1}). The experimental setting is given in Sec.~\ref{exp_setting} followed by the protocol in Sec.~\ref{exp_protocol}.

\subsection{Experimental setting}
\label{exp_setting}
For all Wigner process tomography demonstration experiments, we used a liquid sample of the two-qubit molecule 
chloroform
 dissolved in CD$_3$CN. The $^1$H spin (with a chemical shift of 
 $7.61$ ppm) is defined as the ancilla qubit and the $^{13}$C spin (with a chemical shift of 
 $78.18$ 
ppm) as the system qubit. Thus, the $^1$H and $^{13}$C nuclear spsins of each chloroform molecule 
form a system consisting of two coupled spins 1/2. The coupling constant $J_{HC}$ is $214.15$ Hz. All 
experiments 
 were performed at room temperature (298 K) in a $14.1$ Tesla magnet using a Bruker Avance 
 III 600 spectrometer.

\subsection{Experimental protocol}
\label{exp_protocol}
\begin{figure}
\includegraphics{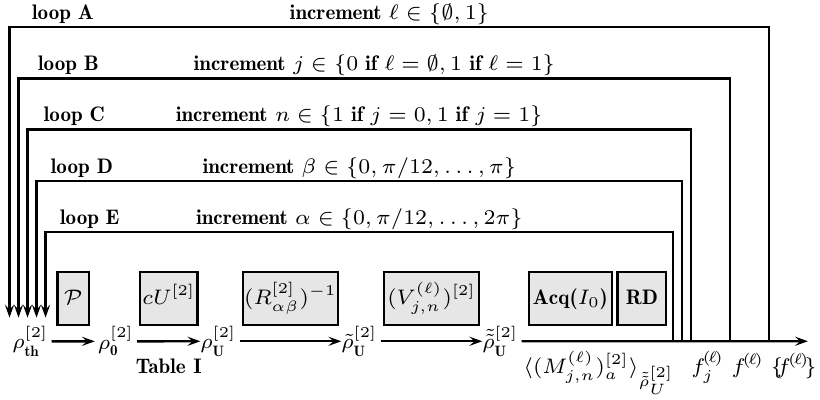}
\caption{Schematic representation of the tomography scheme proposed by Result \ref{eq:prop_reconstr_NMR}
to scan the spherical functions $f^{(\ell)}(\beta,\alpha)$ for the case $N=1$ with $\ell \in \{\emptyset,1\}$ by measuring the expectation values 
$\langle (M^{(\ell)}_{j,n})^{[2]}_a 
\rangle_{\tilde{\tilde{\rho}}^{[2]}_U}$ for $j \in \{0,1\}$, $n=1$, and $a \in \{x,y\}$  as described in Sec.~\ref{caseN1}. See also Fig.~\ref{fig:reconstr}. 
\label{fig:exp} } 
\end{figure}

As summarized in Fig. \ref{fig:exp}, the experiment is composed of six main building blocks. In the first block $\mathcal{P}$, 
the initial operator $\rho_0^{[2]} = 2 I_{0x}^{[2]}$ of the augmented system is prepared from the thermal equilibrium 
density operator. In the high-temperature limit, the deviation density matrix at thermal equilibrium is
proportional to $\rho^{[2]}_{\text{th}} = 2
\sum_{k=0}^{1}\gamma_k I_{kz}$, with $\gamma_k$
being the gyromagnetic ratio of the $k$-th nuclear spin \cite{EBW87, Fahmy2008}.  
The (traceless) operator  $\rho_0^{[2]} = 2 I_{0x}^{[2]}$ is obtained from $\rho^{[2]}_{\text{th}} $
by the following pulse sequence:
\begin{equation}
\mathcal{P} = [\frac{\pi}{2}]_x(I_1) - G - [\frac{\pi}{2}]_y(I_0).
\end{equation}
First, a pulse with flip angle $\frac{\pi}{2}$ and phase $0$ is applied to 
spin $I_1$, followed by a pulsed $B_0$ gradient $G$. This dephases the magnetization of spin $I_1$ and the deviation density operator is proportional to $2 I_{0z}^{[2]}$. Finally, a pulse with flip angle $\pi/2$ and 
phase $\pi/2$ is applied to spin $I_0$, resulting in $\rho_0^{[2]} \propto 2 I_{0x}^{[2]}$.

As described in Sec. \ref{sec:inscribe}, the second block $cU^{[2]}$ realizes 
the controlled propagator based on $U^{[1]}$, which 
transforms $\rho_0^{[2]} =2  I_{0x}^{[2]}$ to $\rho_{U}^{[2]}$. The explicit pulse sequences for 
all used controlled propagators $cU^{[2]}$ are summarized in Table \ref{tab:c_prob}. 

According to Result \ref{eq:prop_reconstr_NMR},  
the third block consists of the rotation  $(R_{\alpha \beta}^{[N+1]})^{-1}$ realized by the pulse $[\beta]_{\alpha-\pi/2}(I_1)$ which transforms the 
operator $\rho_{U}^{[2]}$ into $\tilde{\rho}_{U}^{[2]}$ in order to probe the rank-$j$ components of the
 spherical functions $f_j^{(\ell)}(\beta,\alpha)$ for different polar angles $
\beta$ and azimuthal angles $\alpha$. This rotation is only applied to the system qubit 
$I_1$.

In a general ($N>1$) Wigner process tomography, an additional (fourth) transformation block $(V_{j,n}^{(\ell)})^{[N+1]}$ is required directly after the rotation step (block three) to generate $\tilde{\tilde{\rho}}^{[N+1]}_U$. 
However, as described in Sec.~\ref{caseN1}, for the case $N=1$, we find $(V_{j,1}^{(\ell)})^{[2]} = {\bf 1}^{[2]}$ for all $\ell$ and $j$, which directly results in $\tilde{\tilde{\rho}}^{[2]}_U = \tilde{\rho}^{[2]}_U$.

In the fifth block, the NMR signal of the ancilla 
spin $I_0$ is measured in the acquisition period \textit{Acq} and the expectation values of the four Cartesian operators $I_{0x}^{[2]}$, $I_{0y}^{[2]}$, $(2I_{0x}
I_{1z})^{[2]}$, and $(2I_{0y}I_{1z})^{[2]}$ are obtained as discussed in Sec. \ref{caseN1}. In a 
final step, a relaxation delay \textit{RD} of about $50$ s recovers the initial thermal equilibrium state $
\rho^{[2]}_{\text{th}}$.\\

In a complete tomography experiment, all blocks are repeated multiple times.
The outer loop A cycles through all droplets $\ell \in \{\emptyset,1\}$. Loop B runs over all rank-$j$ components of each droplet $\ell$. Loop C cycles through all Cartesian product operators $(C_{j,n}^{(\ell)})^{[N]}$ appearing in the decomposition
of the axial tensor operator $(T_{j0}^{(\ell)})^{[N]}$ in Eq.~\eqref{eq:trafo_T}. This loop is shown for completeness, although for the considered case of $N=1$, it is obsolete because $n=1$ for all combinations of $\ell$ and $j$.
Finally, the angles $\beta \in \{0, \pi/12, 2\pi/12,\dots, \pi\} $ and $\alpha \in \{0, \pi/12, 2\pi/12,\dots, 
2\pi\} $ are incremented in an equidistant scheme in the two innermost loops D and E (see Fig. \ref{fig:exp}).

\begin{table*}
\caption{Matrix representations for the propagator $U^{[1]}$ and the pulse sequences realizing the corresponding controlled propagator $cU^{[2]}$ [up to a global phase factor (see Appendix \ref{app:globalphase})] to prepare the density matrix $\rho_{U}^{[2]}$ from $\rho_{0}^{[2]}$. $J_{01}$ is the coupling constant between spin $I_0$ and $I_1$. We denote a pulse with flip angle $\beta$ and phase $\alpha$ that is 
applied to spin $k$ by $[\beta]_\alpha(I_k)$. Similarly,
$[\beta_1, \beta_2]_{\alpha_1,\alpha_2}(I_k, I_l)$
specifies two pulses of flip angles $\beta_1$ and $\beta_2$ with phases $\alpha_1$ and $\alpha_2$ that are simultaneously applied to spins $k$ and $l$. 
Here, we use the common shorthand notations $x$, $y$, $-x$, and $-y$ for the phases $0$, $\pi/2$, $\pi$, and $3 \pi/2$, respectively.
\label{tab:c_prob}}
\begin{tabular}{@{\hspace{4mm}} c @{\hspace{7mm}} r @{\hspace{7mm}} l @{\hspace{4mm}}} 
\\[-2mm]
\hline\hline
\\[-3mm]
gate & $U^{[1]}$ \hspace{0.5cm}  & Sequence to prepare $cU^{[2]}$ 
\\[1mm]  \hline \\[-2mm]
Id & $\begin{pmatrix} \hphantom{-} 1 & \hphantom{-} 0\\ \hphantom{-} 0 & \hphantom{-} 1\end{pmatrix}$ & $-$ \vspace{1mm} \\
NOT & $\begin{pmatrix} \hphantom{-} 0 & \hphantom{-} 1\\ \hphantom{-} 1 & \hphantom{-} 0\end{pmatrix}$ & $[\frac{\pi}{2}]_y(I_1)-\frac{1}{2J_{01}}-[\frac{\pi}{2}]_{-y}(I_0,I_1)-[\frac{\pi}{2}]_{-x,x}(I_0,I_1)-[\frac{\pi}{2}]_{y}(I_0)$ \vspace{1mm} \\
$\sqrt{\text{NOT}}$ &  $\frac{1}{2}\begin{pmatrix}  1+{\rm{i}} & 1-{\rm{i}}\\ 1-{\rm{i}} &  1+{\rm{i}}\end{pmatrix}$ & $
[\frac{\pi}{2}]_y(I_1)-\frac{1}{4J_{01}}-[\frac{\pi}{2}]_{-y}(I_0,I_1)-[\frac{\pi}{4}]_{-x,x}(I_0,I_1)-
[\frac{\pi}{2}]_{y}(I_0)$ \vspace{1mm} \\
Hadamard & $\frac{1}{\sqrt{2}}\begin{pmatrix} \hphantom{-} 1 & \hphantom{-} 1\\ \hphantom{-}  1 & -1\end{pmatrix}$ & $[\pi,\frac{\pi}
{4}]_{-x,-y}(I_0,I_1)-\frac{1}{2J_{01}}-[\frac{\pi}{2}]_{x,y}(I_0,I_1)-[\frac{\pi}
{2}]_{y,x}(I_0,I_1)-[\frac{\pi}{2},\frac{\pi}{4}]_{x,-y}(I_0,I_1)$ \vspace{1mm} \\
$\frac{\pi}{2}$ phaseshift & $\frac{1}{\sqrt{2}}\begin{pmatrix} \hphantom{-} 1 & \hphantom{-}  0\\ \hphantom{-} 0 & \hphantom{-} {\rm{i}}\end{pmatrix}$ & 
$[\pi]_x(I_0) -\frac{1}{4J_{01}} - [\pi]_{-x}(I_0)-[\frac{\pi}{2}]_{y}(I_0,I_1)-[\frac{\pi}{4}]_{x}
(I_0,I_1)-[\frac{\pi}{2}]_{-y}(I_0,I_1)$ \vspace{1mm} \\
$\pi$ phaseshift & $\frac{1}{\sqrt{2}}\begin{pmatrix} \hphantom{-} 1 & \hphantom{-} 0\\ \hphantom{-} 0 & -1\end{pmatrix}$ & $
[\pi]_x(I_0) -\frac{1}{2J_{01}} - [\pi]_{-x}(I_0)-[\frac{\pi}{2}]_{y}(I_0,I_1)-[\frac{\pi}{2}]_{x}
(I_0,I_1)-[\frac{\pi}{2}]_{-y}(I_0,I_1)$ \vspace{1mm} \\
$\frac{3\pi}{2}$ phaseshift & $\frac{1}{\sqrt{2}}\begin{pmatrix} \hphantom{-} 1 & \hphantom{-} 0\\ \hphantom{-} 0 & -{\rm{i}}\end{pmatrix}$ 
& $[\pi]_x(I_0) -\frac{3}{4J_{01}} - [\pi]_{-x}(I_0)-[\frac{\pi}{2}]_{y}(I_0,I_1)-[\frac{3\pi}{4}]_{x}
(I_0,I_1)-[\frac{\pi}{2}]_{-y}(I_0,I_1)$ \vspace{1mm} \\
$2\pi$ phaseshift & $\frac{1}{\sqrt{2}}\begin{pmatrix} \hphantom{-} 1 & \hphantom{-} 0\\ \hphantom{-} 0 & \hphantom{-} 1\end{pmatrix}$ & $
[\pi]_x(I_0) -\frac{1}{J_{01}} - [\pi]_{-x}(I_0)-[\frac{\pi}{2}]_{y}(I_0,I_1)-[\pi]_{x}
(I_0,I_1)-[\frac{\pi}{2}]_{-y}(I_0,I_1)$ \vspace{1mm} \\
$[\frac{\pi}{2}]_{x}$ rotation & $\frac{1}{\sqrt{2}}\begin{pmatrix} \hphantom{-} 1 & -{\rm{i}}\\ -{\rm{i}} & 
\hphantom{-} 1\end{pmatrix}$ & $\text{c}[[\frac{\pi}{2}]_x \ \text{rotation}]:=[\frac{\pi}{2}]_{y}(I_1)-\frac{1}{4J_{01}}-[\frac{\pi}{2}]_{-y}(I_1)-
[\frac{\pi}{4}]_{x}(I_1)$ \vspace{1mm} \\
$[\pi]_{x}$ rotation & $\begin{pmatrix} \hphantom{-} 0 & -{\rm{i}}\\ -{\rm{i}} & \hphantom{-} 0\end{pmatrix}$ & $\text{c}[[\pi]_x \ \text{rotation}]:=[\frac{\pi}{2}]_{y}
(I_1)-\frac{1}{2J_{01}}-[\frac{\pi}{2}]_{-y}(I_1)-[\frac{\pi}{2}]_{x}(I_1)$ \vspace{1mm} \\
$[\frac{3\pi}{2}]_{x}$ rotation & $\frac{1}{\sqrt{2}}\begin{pmatrix} -1 & -{\rm{i}}\\ -{\rm{i}} & 
-1\end{pmatrix}$ & c[$[\pi]_{x}$ rotation] $-$ c[$[\frac{\pi}{2}]_{x}$ rotation] \vspace{1mm} \\
$[2\pi]_{x}$ rotation & $\begin{pmatrix} -1 & \hphantom{-} 0\\ \hphantom{-} 0 & -1\end{pmatrix}$ & c[$[\pi]_{x}$ rotation] 
$-$ c[$[\pi]_{x}$ rotation] \vspace{1mm} \\
$[\frac{5\pi}{2}]_{x}$ rotation & $\frac{1}{\sqrt{2}}\begin{pmatrix} -1 & \hphantom{-} {\rm{i}}\\ \hphantom{-} {\rm{i}} & 
-1\end{pmatrix}$ & c[$[\pi]_{x}$ rotation] $-$ c[$[\pi]_{x}$ rotation] $-$ c[$[\frac{\pi}{2}]_{x}$ 
rotation] \vspace{1mm} \\
$[3\pi]_{x}$ rotation & $\begin{pmatrix}  \hphantom{-} 0 & \hphantom{-} {\rm{i}}\\ \hphantom{-} {\rm{i}} &  \hphantom{-} 0\end{pmatrix}$ & c[$[\pi]_{x}$ rotation] $-$ 
c[$[\pi]_{x}$ rotation] $-$ c[$[\pi]_{x}$ rotation] \vspace{1mm} \\
$[\frac{7\pi}{2}]_{x}$ rotation & $\frac{1}{\sqrt{2}}\begin{pmatrix} \hphantom{-} 1 & \hphantom{-} {\rm{i}}\\ \hphantom{-} {\rm{i}} & \hphantom{-} 1\end{pmatrix}
$ & c[$[\pi]_{x}$ rotation] $-$ c[$[\pi]_{x}$ rotation] $-$ c[$[\pi]_{x}$ rotation]  $-$ c[$[\frac{\pi}
{2}]_{x}$ rotation] \vspace{1mm} \\
$[4\pi]_{x}$ rotation & $\begin{pmatrix} \hphantom{-} 1 & \hphantom{-} 0\\ \hphantom{-} 0 & \hphantom{-} 1\end{pmatrix}$ & c[$[\pi]_{x}$ rotation] $-$ 
c[$[\pi]_{x}$ rotation] $-$ c[$[\pi]_{x}$ rotation] $-$ c[$[\pi]_{x}$ rotation] 
\\[1mm]  \hline \hline  \\[-2mm]
\end{tabular}
\end{table*}

\section{Results of NMR experiments}
\label{chapt:results}
Here we present the experimental results of the Wigner process tomography scheme detailed in the previous section for $N=1$. In Sec. \ref{results_idnothad}, we reconstructed the DROPS representations for the Identity, NOT, $\sqrt{ \text{NOT}}$, and Hadamard gates. In Sec. 
\ref{results_phaseshift}, the reconstructed Wigner functions of propagators realizing phase-shift gates for 
different phase shifts are presented. In Sec. \ref{results_rot}, tomographic results of the Wigner representation for rotations around the x-axis are shown for rotation angles between $0$ and $4\pi$.

Note that in the standard DROPS representation introduced in \cite{Garon15}, separate spherical
droplet functions $f^{(\ell)}(\theta,\phi)$ are displayed for operator components corresponding to the identity operator (i.e. operator components that do not contain any spin operator and hence $\ell=\{\emptyset\}$), for linear terms involving only a single spin operator (e.g. $\ell=\{1\}$), and for bilinear terms etc. In the present case of a single qubit ($N=1$), only the two functions $f^{\{\emptyset\}}(\theta,\phi)$ and $f^{\{1\}}(\theta,\phi)$ exist and
it is convenient to combine them into a single spherical function 
$f(\theta,\phi)=f^{\{\emptyset\}}(\theta,\phi)+f^{\{1\}}(\theta,\phi)$. This is possible without compromising the bijectivity of the mapping between operators and droplets \cite{Garon15} because the ranks $j$ of $f^{\{\emptyset\}}(\theta,\phi)$ (with $j=0$) and $f^{\{1\}}(\theta,\phi)$ (with $j=1$) are different.
However, as discussed in more detail in Appendix \ref{app:decmop}, for some applications it can be of advantage to separately plot the droplet functions $f^{\{\emptyset\}}(\theta,\phi)$ and $f^{\{1\}}(\theta,\phi)$.
An explicit functional form of the theoretically expected droplet functions for arbitrary rotations is provided in Appendix \ref{app:functionalform}.

\subsection{Identity, NOT, $\sqrt{ \text{NOT}}$, and Hadamard gates}
\label{results_idnothad}
\begin{figure}
\begin{center}
\includegraphics{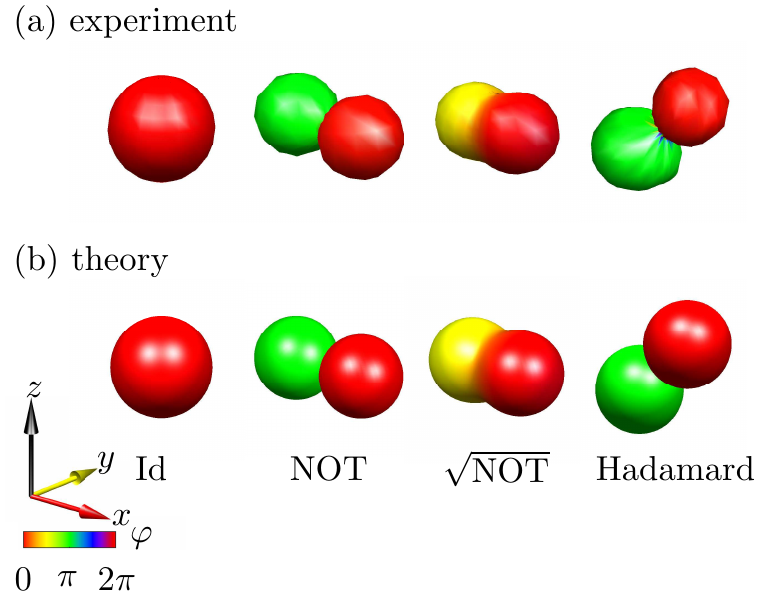}
\caption{(Color online) Experimentally reconstructed (a) and theoretical (b) spherical 
functions $f(\theta,\phi)$ representing propagators for the Id, NOT, $\sqrt{ \text{NOT}}$, and the Hadamard gate.  
The colors red (dark gray), yellow (light gray), green (gray), and blue (black) correspond to phase factors $\exp({\rm{i}} \varphi)$ of 1, i, $-$1, and $-$i \cite{DavidTomo}. See the color bar (grayscale) for $0 \leq \varphi \leq 2\pi$.}
\label{fig:1qubitgates} 
\end{center}
\end{figure}

In an initial series of demonstration experiments, we reconstructed the Wigner functions $f(\theta,\phi)=f^{\{\emptyset\}}(\theta,\phi)+f^{\{1\}}(\theta,\phi)$ for propagators $U^{[1]}$ 
implementing the identity operation Id, a NOT-gate, a square root of a NOT gate ($\sqrt{ \text{NOT}}$), and a Hadamard gate. The matrix representation for 
$U^{[1]}$ and the corresponding pulse sequences for the controlled propagators $cU^{[2]}$ 
are summarized in Table~\ref{tab:c_prob}. The experimental results and the theoretically expected
spherical functions are illustrated in Fig.~\ref{fig:1qubitgates}.  

\subsection{Phase-shift gates}
\label{results_phaseshift}
\begin{figure}
\begin{center}
\includegraphics{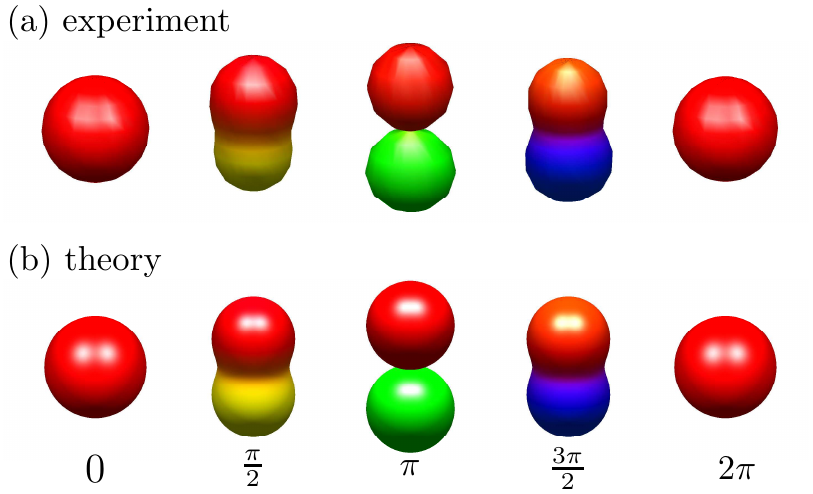}
\caption{(Color online) 
Experimentally reconstructed (a) and theoretical (b) spherical 
functions $f(\theta,\phi)$ representing propagators of phase-shift 
gates for $0$, $\pi/2$, $\pi$, $3\pi/2$, and $2\pi$. 
\label{fig:phaseshift} } 
\end{center}
\end{figure}

We also demonstrate our reconstruction approach to tomograph the Wigner representation $f(\theta,\phi)=f^{\{\emptyset\}}(\theta,\phi)+f^{\{1\}}(\theta,\phi)$ of propagators realizing phase shifts of $0, \pi/2$, $\pi$, $3\pi/2$, and $2\pi$. The matrix representations 
for these gates are shown in Table~\ref{tab:c_prob} and the results of the tomographic 
reconstruction for the spherical functions are presented in Fig.~\ref{fig:phaseshift}.
In this figure, we also show the results for the phase-shift gate corresponding
to a phase shift of $0$, for which the pulse sequence is identical to cId (see Table \ref{tab:c_prob}).
This figure reflects the $2\pi$ periodicity of the phase-shift gates. The propagator representing a phase shift of 0 is identical to the identity operator $U^{[N]}(0)={\bf 1}^{[N]}$ which is a positive red (dark gray) sphere in the DROPS representation. 
After a phase shift of $2\pi$, the propagator becomes the identity operator again, i.e., $U^{[N]}(2\pi)=U^{[N]}(0)={\bf 1}^{[N]}$, which is represented by the same red (dark gray) sphere.
This highlights the contrast to the spinor property of propagators for rotations of a spin-1/2 particle 
as shown in Sec.~\ref{results_rot} which results in a periodicity of $4\pi$ (see Appendix \ref{app:periodicity}).

\subsection{Rotations}
\label{results_rot}
\begin{figure}
\begin{center}
\includegraphics[width=1\linewidth]{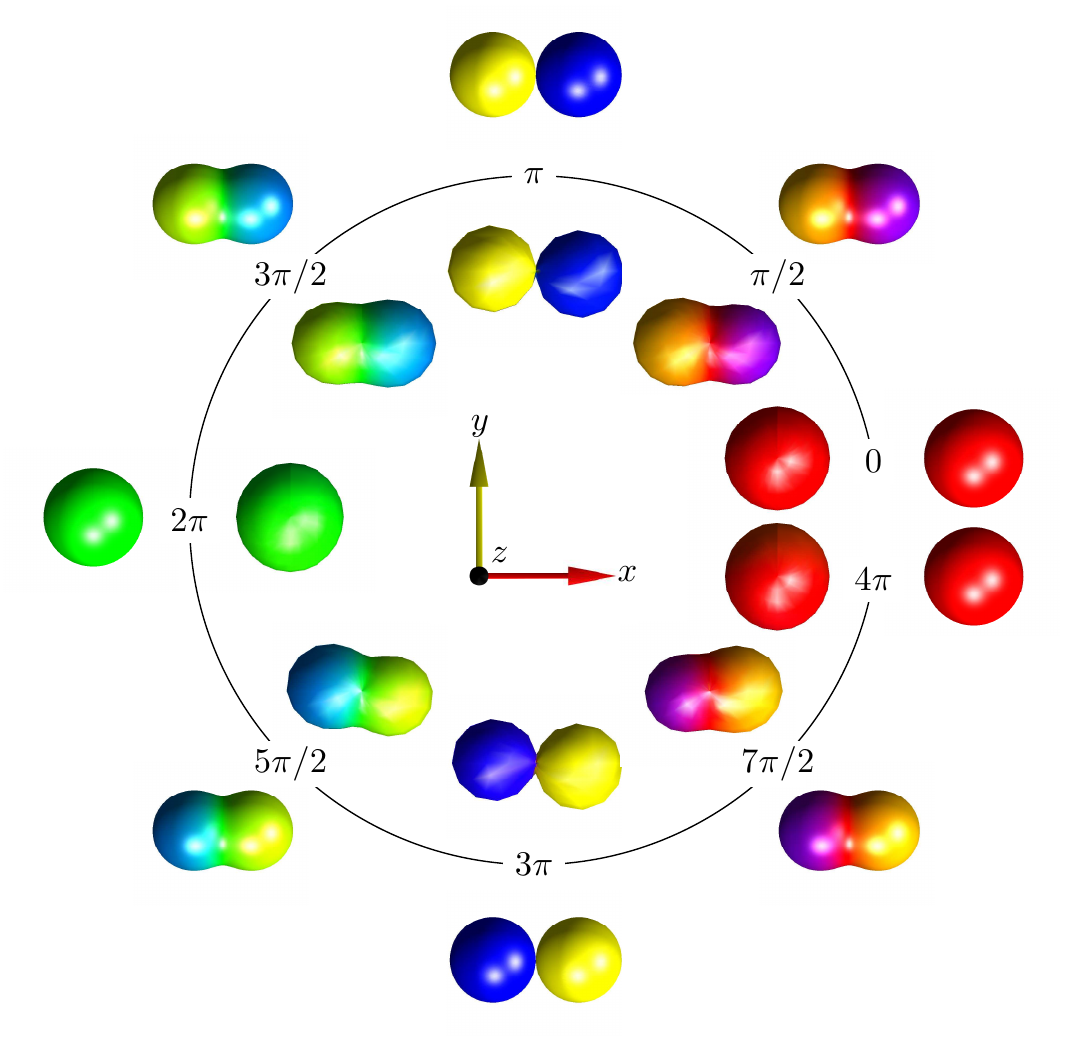}
\caption{(Color online) Reconstructed spherical functions $f(\theta,\phi)$ representing propagators for 
rotations  
around the $x$-axis for  rotation angles $0, \pi/2, \pi, 3\pi/2, 2\pi, 5\pi/2, 3\pi, 7\pi/2$, and $4\pi$. 
The experimentally sampled shapes are positioned along the inner circle, whereas the 
theoretical functions are positioned along the outer circle.
The colors red (dark gray), yellow (light gray), green (gray), and blue (black) correspond to phase factors $\exp({\rm{i}} \varphi)$ of 1, i, $-$1, and $-$i; see the color bar (gray scale) in Fig. \ref{fig:1qubitgates}.
Note the sign change of the experimentally measured and theoretical DROPS representation when a given rotation angle 
is increased by 2$\pi$, nicely illustrating the spinor property of propagators corresponding to rotations of spin-1/2 particles.
 \label{fig:rotation} } 
\end{center}
\end{figure}

We also reconstructed the Wigner functions $f(\theta,\phi)=f^{\{\emptyset\}}(\theta,\phi)+f^{(1)}\{\theta,\phi\}$ of propagators corresponding to rotations 
around the $x$ axis 
for 
rotation angles $0, \pi/2, \pi, 3\pi/2, 2\pi, 5\pi/2, 3\pi, 7\pi/2$, and $4\pi$. 
Again, the matrix representations for the given propagators are summarized in Table~
\ref{tab:c_prob} and the tomography results are shown in Fig.~\ref{fig:rotation}.

This figure probably represents one of the most direct and compelling visualizations
of the experimentally measured spinor property of the propagators corresponding to the rotation of a spin-1/2 particle.
For a rotation angle of 0, the propagator $U^{[N]}(0)$ corresponds to the identity operator ${\bf 1}^{[N]}$, which is
represented by a positive red (dark gray) sphere in the DROPS representation.
However, for  a rotation angle of $2\pi$, the propagator $U^{[N]}(2\pi)$ does not correspond to the identity operator ${\bf 1}^{[N]}$ but to $-{\bf 1}^{[N]}$, represented
by a negative green (gray) sphere in the DROPS representation, i.e., $U^{[N]}(2\pi)=-U^{[N]}(0)$. Only after a rotation of $4\pi$, the propagator becomes the identity operator again, i.e., $U^{[N]}(4\pi)=U^{[N]}(0)={\bf 1}^{[N]}$ 
\cite{Thompson1994} (see Appendix \ref{app:periodicity}).

\section{Discussion}
\label{sec:discussion}
Here, we introduced a general approach to experimentally measure the Wigner representation of quantum-mechanical propagators. The approach based on Result \ref{eq:prop_reconstr} can be applied to individual quantum systems consisting of qubits studied in the context of quantum information processing. Result \ref{eq:prop_reconstr_NMR} extends this approach to ensembles of spin systems studied in nuclear magnetic resonance (NMR) or electron spin resonance (ESR). Demonstration experiments have been implemented in an NMR setting.

A useful property of the DROPS Wigner representation of rotations is the fact that the orientation of the droplet shape
directly reflects the orientation of the rotation axis. For example, for the case of rotations around the $x$ axis [see Fig. \ref{fig:rotation}] the droplet shapes are aligned along the $x$ axis, whereas in the case of the Hadamard gate, the rotation axis is aligned along the bisector of the angle between the $x$ and the $z$ axis [see Fig. \ref{fig:1qubitgates}].
Furthermore, the DROPS Wigner representation makes it possible to directly see the spinor property of rotation propagators, which results in a sign change that is reflected by a corresponding color change. For example, for rotation angles of 0 and 2$\pi$, the propagators are represented by a red and and a green sphere (dark gray and gray), respectively. For a rotation angle of $\pi$, the droplet consists of a blue and yellow sphere (black and light gray) and for a rotation angle of $3\pi$ the colors are interchanged.

A reasonable match between the experimentally reconstructed and 
theoretical predicted spherical functions is found in Figs.~\ref{fig:1qubitgates}-\ref{fig:decomp_rot_exp}.
Deviations are due to experimental imperfections, such as 
finite experimental signal-to-noise ratio, finite accuracy of pulse calibration, 
$B_0$ and $B_1$ inhomogeneities~\cite{levitt_book, EBW87}, 
pulse shape distortions due to the amplifiers and the finite bandwidth of the resonator~\cite{transient},  
relaxation losses during the preparation block, the implementation of the $cU$ operation and the detection block,
partial saturation of the signal due to a finite relaxation period between scans,
radiation damping effects~\cite{CMR:CMR1}, and
truncation effects in the automated integration and comparison of the spectra.

As discussed previously in \cite{DavidTomo}, the Wigner tomography can only measure the deviation density matrix, i.e., the
{\it traceless} part of a given {\it density} operator $\rho^{[N]}$ when performed using standard NMR or ESR
 experiments. 
 This is because experimentally available 
observables in magnetic resonance experiments are traceless operators.
This restriction is irrelevant in most cases of practical interest, because the part of the 
density operator that is proportional to the identity operator neither evolves during an experiment nor 
gives rise to any detectable signals in the experiments.
However for {\it propagators}, the part that is proportional to the identity operator plays an important role and 
it would be a serious restriction if only the traceless part of a propagator $U^{[N]}$ could be 
determined using NMR or ESR experiments. Fortunately, this is not the case for the presented 
approach  based on the mapping of the propagator $U^{[N]}$ to {\it off-diagonal} blocks of 
the density operator $\rho^{[N+1]}$ of the spin system augmented by an additional ancilla spin, i.e. also in 
standard NMR and ESR experiments the {\it full} propagator $U^{[N]}$ can be reconstructed and not 
only its traceless part. 
For example, for a system consisting only of a single spin 1/2, this allows us to distinguish 
the propagators 
corresponding to rotations about the same axis but with different rotation angles
$\pi - \alpha$ and $\pi + \alpha$, which for arbitrary angles $\alpha$ 
only differ by the sign of the identity operator contribution to $U^{[N]}$. This is illustrated in Figs. \ref{fig:rotation}, \ref{fig:decomp_rot_sim}, and \ref{fig:decomp_rot_exp},
where due to the sign of the contribution of the identity operator, e.g., the propagators of $\pi/2$ and $3 \pi/2$ can be clearly distinguished.

The presented proof-of-principle experiments exploited the fact that
a controlled propagator $cU^{[N+1]}$ can be constructed in a straight forward way for any given (i.e. previously known) unitary operator 
$U^{[N]}$.
However, it is important to note that this tomography approach based on a single ancilla qubit can be generalized to the experimental reconstruction of  the Wigner function representing an {\it unknown} propagator
$U^{[N]}$
based on additional ancilla qubits and controlled-SWAP operations \cite{Zhou2011}.
As discussed in more detail in Appendix \ref{app:errorterms},
in this case, the Wigner process tomography can provide valuable information to identify sources of systematic and random errors
of a given implementation of a desired unitary transformation.

The objective of this work was to show how the
spherical functions of the DROPS representation of a set of propagators of interest can be experimentally scanned. 
If no {\it a priori} information of the expected droplet shapes is available,
a large number of sampling points is necessary to ensure that all features of the shapes
are captured in sufficient detail.
Here, the experimental scanning procedure was demonstrated
using a simple sampling scheme with equidistant steps of the polar and azimuthal angles $\beta$ and $\alpha$ as shown in Fig.~\ref{fig:reconstr}. 
Based on previous simulations, increments of $\pi/12$ were chosen for $\beta$ and $\alpha$ in the presented demonstration experiments, which formed a reasonable compromise between the overall duration of the experiments and 
a sufficiently good resolution to visually show all characteristic features of the droplet functions.
However, as previously discussed in the context of Wigner tomography of density operators  \cite{DavidTomo},
more efficient sampling schemes are available.

In the ideal case of noiseless data, only a relatively small number of sampling points would be necessary to 
determine the correct expansion coefficients of spherical harmonics.
This is a result of the fact that each droplet function is band-limited; i.e., for each function $f^{(\ell)}$ the maximum value $j_{max}^{(\ell)}$  of the rank $j\in J(\ell)$ is known (see Eq. \eqref{Ajb})
\cite{Garon15}.
For example, in the case $N=2$ shown in Fig. \ref{fig:drops}, the maximal ranks of the different droplet functions are
$j_{max}^{\{ \emptyset \}}=0$, $j_{max}^{\{ 1 \}}=j_{max}^{\{ 2 \}}=1$ and $j_{max}^{\{ 12 \}}=2$.
The minimal number of sampling points for band-limited spherical functions is given by the {\it optimal dimensionality} $(j_{max}^{(\ell)}+1)^2$ \cite{Khalid2014} and a number of different sampling strategies have been 
proposed in the literature which approach this limit. The equiangular schemes based the sampling theorems by Driscoll \& Healy \cite{Driscoll1994, Kennedy2013} 
and  McEwen \& Wiaux \cite{McEwen2011} exceed the minimal number of sampling points by factors of four and two, respectively.
The Gauss-Legendre quadrature on the sphere  \cite{Shukowsky1986} requires $(2 j_{max}^{(\ell)}+1) (j_{max}^{(\ell)}+1)$ samples.
Recently, an optimal iso-latitude sampling scheme with a fast and stable spherical harmonics transform was proposed by Khalid et al. \cite{Khalid2014}. 
In the limit of large $j_{max}^{(\ell)}$, the optimal dimensionality is also approached by Lebedev {\it two-angle sets} \cite{leb75,leb76,leb99}.
Related Lebedev {\it three-angle} sets have been previously used for the experimental decomposition of detected NMR signals in terms of the rank $j$ and order $m$ of the {\it density operator} at a chosen time point during an experiment \cite{Levitt2005}. In the presence of noise and instrumental variations it was found that Lebedev sets with large differences in the weights provide larger errors compared to sets with nearly uniform weights \cite{Levitt2005}.

It is interesting to note that for some droplet functions $f^{(\ell)}$ the number of sampling points may be even further reduced compared to the 
optimal dimensionality  $(j_{max}^{(\ell)}+1)^2$
of band-limited spherical functions because  the set $J(\ell)$ of ranks $j$ in general does not include all values of $j$ between 0 and $j_{max}^{(\ell)}$
\cite{Garon15}. For example,  the set $J(\ell)$ of possible $j$ values only consists of $j =j_{max}^{\{ 1 \}} =1$ 
(and does not include the case of $j=0$) for the droplet function $f^{\{ 1 \}}$, see Sec. \ref{caseN1} and Figs. 
\ref{fig:decomp} - \ref{fig:decomp_rot_exp}. Furthermore, for the analysis of some error terms, only a subset of all droplet functions may be of interest, which makes it possible to further decrease the number of measurements.

\section{Conclusion}
\label{sec:conclusion}
In this work, we have theoretically developed and experimentally demonstrated a 
Wigner process tomography scheme by extending the reconstruction approach of the 
Wigner representation for multi-qubit states of \cite{DavidTomo}.
Our scheme reconstructs the relevant spherical functions representing propagators by
imprinting these operators on the density matrix of an augmented system with an additional
ancilla spin and subsequently measuring expectation values of rotated axial tensor operators.
The approach is universally applicable and 
not restricted to NMR methodologies or to particles with spin $1/2$.
In the presented proof-of-concept experiments, a reasonable match was found between the theoretical and the experimentally reconstructed
Wigner functions of a range of important single-qubit propagators. 
In particular, the method provided a direct demonstration of the  spinor property of propagators corresponding to the rotation of a spin 1/2 particle.

\section{Acknowledgements}
D.L. acknowledges support from the PhD program
{\it Exploring Quantum Matter} (ExQM) within the
{\it Excellence Network
of Bavaria} (ENB). S.J.G.
acknowledges support from the Deutsche Forschungsgemeinschaft
(DFG) through Grant No. Gl 203/7-2. We
thank Raimund Marx for providing the NMR samples. The
experiments were performed at the Bavarian NMR Center (BNMRZ)
at the Technical University of Munich.

\section{Appendix}

\subsection{Example of the imprinting of a propagator $U^{[N]}$ on the density operator $\rho_U^{[N+1]}$ of an augmented spin system}
\label{append_example_cU}
Here the operators $U^{[N]}$, $cU^{[N+1]}$, $\rho_0^{[N+1]}$ and $\rho_U^{[N
+1]}$ defined in Sec. \ref{sec:inscribe} of the main text are explicitly given for a simple example,
where the {\it system of interest} consists only of one ($N$=1) spin 1/2 denoted $I_1$, 
which is augmented by an {\it ancilla} spin 1/2 denoted $I_0$. In this case, the matrix representation of the 
propagator $U^{[N]}=U^{[1]}$ to be scanned is of dimension $2\times 2$ and has the general 
form 
\begin{equation}\label{Usimp}
U^{[1]}= \begin{pmatrix} u_{11} & u_{12} \\ u_{21}  & u_{22}  \end{pmatrix}.
\end{equation}
The matrix representation of the controlled propagator $cU^{[N+1]}=cU^{[2]}$ in the augmented 
system is of dimension $4\times 4$ and is given by
\begin{equation}\label{cUsimp}
cU^{[2]}= 
\begin{pmatrix} 
1 & 0 & 0 & 0\\
0 & 1 & 0 & 0\\
0 & 0 & u_{11} & u_{12} \\ 
0 & 0 & u_{21}  & u_{22}  \end{pmatrix}.
\end{equation}
For the prepared initial state of the augmented spin system, the deviation
density operator [see Eq. \eqref{rh}] is proportional to
\begin{equation}\label{rhoisimp}
\rho_0^{[2]}= 2 I_{0x}^{[N+1]}=
\begin{pmatrix} 
0 & 1\\
1 & 0\end{pmatrix}
\otimes
\begin{pmatrix} 
1 & 0\\
0 & 1\end{pmatrix}
=
\begin{pmatrix} 
0 & 0 & 1 & 0\\
0 & 0 & 0 & 1\\
1 & 0 & 0 & 0 \\ 
0 & 1 & 0 & 0  \end{pmatrix}
\end{equation}
and applying Eq. (\ref{rhoU}) results in
\begin{equation}\label{rhoUsimp}
\rho_U^{[2]}= cU^{[2]} \ \rho_0^{[2]} \ (cU^{[2]})^\dagger
=
\begin{pmatrix} 
0 & 0 & u^\ast_{11} & u^\ast_{21}\\
0 & 0 & u^\ast_{12}&u ^\ast_{22}\\
u_{11} & u_{12} & 0 & 0 \\ 
u_{21} & u_{22} & 0 & 0  \end{pmatrix}.
\end{equation}
Note that this operator is still Hermitian but according to Eq. (\ref{rhoU1}) can be 
decomposed
into the two non-Hermitian operators $I^- \otimes U^{[1]}$ and $I^+ \otimes U^{[1]}$:
\begin{eqnarray}\label{rhoUsimpdecomp}
\rho_U^{[2]}&=& 
\begin{pmatrix} 
0 & 0\\
1 & 0\end{pmatrix}
\hspace{-1mm}
\otimes 
\hspace{-1mm}
\begin{pmatrix} 
u_{11} & u_{12} \\ 
u_{21}  & u_{22}  
\end{pmatrix}
{+}
\begin{pmatrix} 
0 & 1\\
0 & 0\end{pmatrix}
\hspace{-1mm}
\otimes
\hspace{-1mm}
\begin{pmatrix} 
u^\ast_{11} & u^\ast_{21}\\
u^\ast_{12}&u ^\ast_{22}
\end{pmatrix}\\
&=&
\hskip 0.5cm \begin{pmatrix} 
0 & 0 & 0 & 0\\
0 & 0 & 0 & 0\\
u_{11} & u_{12} & 0 & 0 \\ 
u_{21} & u_{22} & 0 & 0  \end{pmatrix}
+
\begin{pmatrix} 
0 & 0 & u^\ast_{11} & u^\ast_{21}\\
0 & 0 & u^\ast_{12}&u ^\ast_{22}\\
0 & 0 & 0 & 0 \\ 
0 & 0 & 0 & 0  \end{pmatrix}. \nonumber
\end{eqnarray}

\subsection{Proof of Result 1}
\label{proof}
We start with Eq. \eqref{tomoprime} (see also Result 2 in \cite{DavidTomo}).
If the density operator of an augmented 
spin system, consisting of an ancilla spin $I_0$ in addition to the spins $I_1, ..., I_N$,
can be prepared in the state 
\begin{equation}
\rho^{[N+1]}_A=\begin{pmatrix} {\bf 0}^{[N]} & 
(A^{[N]})^{\dag} \\ A^{[N]} & {\bf 0}^{[N]} \end{pmatrix},
\end{equation}
the (in general complex) value of the scalar product $\langle {T}_{j, \alpha \beta}^{(\ell)[N]} | 
A^{[N]}\rangle$ can be experimentally obtained by measuring the expectation value
of the operator 
$
I^+ \otimes {T}_{j, \alpha \beta}^{(\ell)[N]}
$:
\begin{eqnarray} 
\label{expra}
\left \langle  I^+ \hspace{-1mm} \otimes \hspace{-1mm} {T}_{j, \alpha \beta}^{(\ell)[N]} \right \rangle_{\hspace{-1mm} \rho^{[N
{+}1]}_A} 
\hspace{-1mm}&=&{\rm tr} \left \{[I^+ \hspace{-1mm} \otimes  {T}_{j, \alpha \beta}^{(\ell)[N]} ] \rho^{[N+1]}
_A \right \} \\
\hspace{-1mm} &=&
  {\rm tr} \left \{\begin{pmatrix} 
0 & {T}_{j, \alpha \beta}^{(\ell)[N]}  \nonumber \\
0 & 0\end{pmatrix} \hspace{-2mm}
\begin{pmatrix} {\bf 0}^{[N]} & (A^{[N]})^{\dag} \\ A^{[N]} & {\bf 0}^{[N]} \end{pmatrix}
\hspace{-1mm} \right \} \nonumber  \\
&=&
  {\rm tr} \left \{\begin{pmatrix} 
 {T}_{j, \alpha \beta}^{(\ell)[N]} A^{[N]} &    {\bf 0}^{[N]}   \nonumber  \\
{\bf 0}^{[N]} & {\bf 0}^{[N]} \end{pmatrix}
\right \}\\
&=&
  {\rm tr} \left \{
 {T}_{j, \alpha \beta}^{(\ell)[N]} A^{[N]} \right \}  \nonumber  \\
 &=&
  {\rm tr} \left \{
 ({T}_{j, \alpha \beta}^{(\ell)[N]})^\dagger A^{[N]} \right \}  \nonumber \\
 &=& \left \langle {T}_{j, \alpha \beta}^{(\ell)[N]} | A^{[N]} \right \rangle. \nonumber
\end{eqnarray}
Note that the operator $I^+ \otimes {T}_{j, \alpha \beta}^{(\ell)[N]}$ is not Hermitian and hence is 
not an observable that can be directly measured.
However, based on the definition of $I^+$ \cite{EBW87}, we can decompose the  expectation value of Eq. 
(\ref{expra}) into a complex linear combination of the expectation values of the Hermitian 
operators 
$I_x  \otimes {T}_{j, \alpha \beta}^{(\ell)[N]}$ and $I_y  \otimes {T}_{j, \alpha \beta}^{(\ell)[N]}$:
\begin{align}
\langle  I^+ \hspace{-1mm} \otimes {T}_{j, \alpha \beta}^{(\ell)[N]} & \rangle_{\hspace{-1mm}\rho^{[N
+1]}_A} 
\hspace{-1mm} \\ \nonumber 
&=\left \langle  (I_x + {i} I_y) \otimes {T}_{j, \alpha \beta}^{(\ell)[N]} 
\right \rangle_{\rho^{[N+1]}_A} \\
&=\left \langle  I_x  \hspace{-1mm} \otimes {T}_{j, \alpha \beta}^{(\ell)[N]} \right \rangle_{\rho^{[N+1]}_A} 
{+} {i} \ \left \langle  I_y \hspace{-1mm} \otimes {T}^{[N]}_{\alpha \beta} 
\right \rangle_{\hspace{-1mm}\rho^{[N+1]}_A}.  \nonumber 
\end{align}

\subsection{Example of the measurement of the scalar product between a rotated axial tensor operator and a propagator}
\label{append_example_meas}
Considering the same spin system as in Appendix~\ref{append_example_cU}, we show how the procedure outlined in the main text allows us to obtain
scalar products of the form [see Eq.~\eqref{eq:scalU}]
\begin{align}
\label{expected}
\langle &{T}_{j, \alpha \beta}^{(\ell)[N]} | U^{[N]}\rangle \\
&= \langle  I_x  \otimes {T}_{j, \alpha \beta}^{(\ell)[N]} \rangle_{\rho^{[N+1]}_U} 
+ {i} \ \langle  I_y  \otimes {T}_{j, \alpha \beta}^{(\ell)[N]} \rangle_{\rho^{[N+1]}_U}. \nonumber
\end{align}
Here, the operator 
\begin{equation}
{T}_{j, \alpha \beta}^{(\ell)[N]}=R^{[N]}_{\alpha\beta} (T_{j0}^{(\ell)})^{[N]} 
(R^{[N]}_{\alpha\beta})^\dagger
\end{equation}
is a rotated axial tensor operator $(T_{j0}^{(\ell)})^{[N]}$.

For example, the rank $j=1$ axial tensor operator for the droplet corresponding to spin $I_1$
is given by 
\begin{equation}
(T_{10}^{(1)})^{[1]}
=\sqrt{2} I_{1z}^{[1]}
=
\begin{pmatrix} 
1/\sqrt{2} & 0\\
0&  -1/\sqrt{2}
\end{pmatrix}
\end{equation}
and for $\alpha=0$ and $\beta=\pi/2$, the rotated axial tensor operator is
\begin{align}
{T}_{1, 0, \pi/2}^{(1)[1]}&=R^{[1]}_{0,\pi/2} (T_{10}^{(1)})^{[N]} 
(R^{[1]}_{0,\pi/2})^\dagger \\ \nonumber
&=\sqrt{2} I_{1x}^{[1]}
=
\begin{pmatrix} 
0& 1/\sqrt{2} \\
1/\sqrt{2} & 0
\end{pmatrix}.
\end{align}
With
\begin{align}
I_x  \otimes {T}_{1, 0, \pi/2}^{(1)[1]}&=
\begin{pmatrix} 
0 & 1/2\\
1/2 & 0\end{pmatrix}
\otimes
\begin{pmatrix} 
0& 1/\sqrt{2} \\
1/\sqrt{2} & 0
\end{pmatrix} \\
&=
\frac{1}{2\sqrt{2}}
\begin{pmatrix} 
\ \ 0 &\ \ 0 &\ \ 0 &\ \ 1\\
\ \ 0 &\ \ 0 &\ \ 1 &\ \ 0\\
\ \ 0 &\ \ 1 &\ \ 0 &\ \ 0 \\ 
\ \ 1 &\ \ 0 &\ \ 0 &\ \ 0  \end{pmatrix} \nonumber
\end{align}
and
\begin{align}
I_y  \otimes {T}_{1, 0, \pi/2}^{(1)[1]}&= 
\begin{pmatrix} 
0 & -{i}/2\\
{i}/2 & 0\end{pmatrix}
\otimes
\begin{pmatrix} 
0& 1/\sqrt{2} \\
1/\sqrt{2} & 0
\end{pmatrix}  \\
&=
-\frac{i}{2\sqrt{2}}
\begin{pmatrix} 
\ \ 0 & \ \ 0 & \ \ 0 & \ \ 1 \\
\ \ 0 & \ \ 0 & \ \  1  &\ \  0\\
\ \ 0 & -1 &\ \  0 &\ \  0 \\ 
-1 & \ \ 0 &\ \  0 &\ \  0  \end{pmatrix}  \nonumber
\end{align}
we find the expectation values 
\begin{align}
& \langle  I_x  \otimes {T}_{1, 0, \pi/2}^{(1)[1]} \rangle_{\rho^{[2]}_U} \\ 
&=
\frac{1}{2\sqrt{2}} \ {\rm tr} \left \{
\begin{pmatrix} 
\ \ 0 &\ \ 0 &\ \ 0 &\ \ 1\\
\ \ 0 &\ \ 0 &\ \ 1 &\ \ 0\\
\ \ 0 &\ \ 1 &\ \ 0 &\ \ 0 \\ 
\ \ 1 &\ \ 0 &\ \ 0 &\ \ 0  \end{pmatrix}
\begin{pmatrix} 
0 & 0 & u^\ast_{11} & u^\ast_{21}\\
0 & 0 & u^\ast_{12}&u ^\ast_{22}\\
u_{11} & u_{12} & 0 & 0 \\ 
u_{21} & u_{22} & 0 & 0  \end{pmatrix}
   \right  \} \nonumber \\
&=
 \frac{1}{2\sqrt{2}} \ {\rm tr} \left \{
\begin{pmatrix} 
u_{21} & 0 & 0 & 0\\
0 & u_{12} & 0&0\\
0 & 0 & u^\ast_{12} & 0 \\ 
0 & 0& 0 & u^\ast_{21}  \end{pmatrix} \right \} \nonumber \\
&=
\frac{1}{2\sqrt{2}} \ (u_{21} + u_{12}  + u^\ast_{12} +u^\ast_{21})\nonumber \\
&=
\frac{1}{\sqrt{2}}\  ({\rm Re}\{ u_{12}\} + {\rm Re}\{ u_{21}\} ) \nonumber
\end{align}
and
\begin{align} 
&\langle  I_y  \otimes {T}_{1, 0, \pi/2}^{(1)[1]} \rangle_{\rho^{[2]}_U} 
\\
&=
-\frac{i}{2\sqrt{2}} \ {\rm tr} \left \{
\begin{pmatrix} 
\ \ 0 & \ \ 0 & \ \ 0 & \ \ 1 \\
\ \ 0 & \ \ 0 & \ \  1  &\ \  0\\
\ \ 0 & -1 &\ \  0 &\ \  0 \\ 
-1 & \ \ 0 &\ \  0 &\ \  0  \end{pmatrix}
\begin{pmatrix} 
0 & 0 & u^\ast_{11} & u^\ast_{21}\\
0 & 0 & u^\ast_{12}&u ^\ast_{22}\\
u_{11} & u_{12} & 0 & 0 \\ 
u_{21} & u_{22} & 0 & 0  \end{pmatrix}  
    \right \} \nonumber \\
&=
-\frac{i}{2\sqrt{2}} \ {\rm tr}  \left \{
\begin{pmatrix} 
u_{21} & 0 & 0 & 0\\
0 & u_{12} & 0&0\\
0 & 0 & -u^\ast_{12} & 0 \\ 
0 & 0& 0 & -u^\ast_{21}  \end{pmatrix} \right \}  \nonumber \\
&=
-\frac{i}{2\sqrt{2}} (u_{21}  + u_{12}  - u^\ast_{12} -u^\ast_{21}) \nonumber \\
&=
\frac{1}{\sqrt{2}}\  ({\rm Im}\{ u_{12}\} + {\rm Im}\{ u_{21}\} ). \nonumber
\end{align} 
Hence, we find
\begin{align} 
&\langle  I_x  \otimes {T}_{1, 0, \pi/2}^{(1)[1]} \rangle_{\rho^{[N+1]}_U} 
{+} {i} \ \langle  I_y  \otimes {T}_{1, 0, \pi/2}^{(1)[1]} \rangle_{\rho^{[N+1]}_U} \\
&= \frac{1}{\sqrt{2}}\  ({\rm Re}\{ u_{12}\}{+}  {i} \ {\rm Im}\{ u_{12}\}{+} {\rm Re}\{ u_{21}\}{+}  {i} \ {\rm Im}\{ u_{21}\}\} )\hskip 0.7cm \nonumber \\
&=\frac{1}{\sqrt{2}}\  (u_{12} {+}u_{21} ). \nonumber
\end{align}

This is identical to the scalar product
\begin{align}  
\langle {T}_{1, 0, \pi/2}^{(1)[1]}| U^{[1]}\rangle=&  
{\rm tr} \{
({T}_{1, 0, \pi/2}^{(1)[1]})^\dagger U^{[1]}
\}\\
=&
{\rm tr} \left \{
\begin{pmatrix} 
0& 1/\sqrt{2} \\
1/\sqrt{2} & 0
\end{pmatrix}^\dagger
\begin{pmatrix} u_{11} & u_{12} \\ u_{21}  & u_{22}  \end{pmatrix}
 \right \} \nonumber \\
=&
{\rm tr} \left \{
\begin{pmatrix} 
0& 1/\sqrt{2} \\
1/\sqrt{2} & 0
\end{pmatrix}
\begin{pmatrix} u_{11} & u_{12} \\ u_{21}  & u_{22}  \end{pmatrix}
 \right \} \nonumber \\
=&1/\sqrt{2} \ 
{\rm tr} \left \{
\begin{pmatrix} 
u_{21}& u_{22} \\
u_{11} & u_{12}
\end{pmatrix}
\right \} \nonumber \\
=&1/\sqrt{2} \ 
(u_{12}+u_{21}), \nonumber
\end{align}  
see Eq. \eqref{expected}.

\subsection{Proof of Equations \eqref{eq:rot_exp_1}-\eqref{eq:rot_exp_2}}
\label{app:proof_inv_rot}
Here we show that
\begin{align}
\langle  I_a  {\otimes}  {T}_{j, \alpha \beta}^{(\ell)[N]}
 \rangle_{\rho^{[N+1]}_U} &= \langle  I_a  
{\otimes} (T_{j0}^{(\ell)})^{[N]}  \rangle_{\tilde{\rho}^{[N+1]}_U}
\end{align}
for $a \in \{x,y\}$. This can be seen by a sequence of reformulations and by exploiting properties of the trace operation and of the tensor product:
\begin{align}
&\langle  I_a   {\otimes}  {T}_{j, \alpha \beta}^{(\ell)[N]} 
 \rangle_{\rho^{[N+1]}_U}  \\ \nonumber
&= \tr \{ [I_a \otimes {T}_{j, \alpha \beta}^{(\ell)[N]}] \rho^{[N+1]}_U\} \\ \nonumber
&= \tr \{ [I_a \otimes R^{[N]}_{\alpha\beta} (T_{j0}^{(\ell)})^{[N]} 
(R^{[N]}_{\alpha\beta})^\dagger] \rho^{[N+1]}_U \} \\ \nonumber
&= \tr \{ [ {\bf 1}^{[1]} \otimes R^{[N]}_{\alpha\beta}][I_a \otimes (T_{j0}^{(\ell)})^{[N]}] [ {\bf 1}^{[1]} \otimes R^{[N]}_{\alpha\beta}]^\dagger \rho^{[N+1]}_U \} \\ \nonumber
&= \tr \{ R^{[N+1]}_{\alpha\beta}[I_a \otimes (T_{j0}^{(\ell)})^{[N]}]   (R^{[N+1]}_{\alpha\beta})^\dagger \rho^{[N+1]}_U\} \\\ \nonumber
&= \tr \{ [I_a \otimes (T_{j0}^{(\ell)})^{[N]}]  (R^{[N+1]}_{\alpha\beta})^\dagger \rho^{[N+1]}_U R^{[N+1]}_{\alpha\beta}\} \\ \nonumber
&= \tr \{ [I_a \otimes (T_{j0}^{(\ell)})^{[N]}]  \tilde{\rho}^{[N+1]}_U \}   \\ \nonumber
&= \langle  I_a  
{\otimes} (T_{j0}^{(\ell)})^{[N]}  \rangle_{\tilde{\rho}^{[N+1]}_U} \nonumber
\end{align}
with 
\begin{align}
\tilde{\rho}^{[N+1]}_U &= (R^{[N+1]}_{\alpha\beta})^{\dagger} \rho^{[N+1]}_U R^{[N+1]}_{\alpha\beta} \\ \nonumber
&=(R^{[N+1]}_{\alpha\beta})^{-1} \rho^{[N+1]}_U R^{[N+1]}_{\alpha\beta},
\end{align}
$R^{[N+1]}_{\alpha\beta} = {\bf 1}^{[1]} \otimes R^{[N]}_{\alpha\beta}$ and $a \in \{x,y \}$. Here, we used the properties of the tensor product $AB \otimes CD  = (A \otimes C)(B \otimes D)$ and the invariance of the trace operation under cyclic permutations. Note that the rotation $R^{[N+1]}_{\alpha\beta}$ affects only the system qubits $I_1, \dots ,I_N$ and not the ancilla qubit $I_0$.

\subsection{Role of global phase factors in $U^{[N]}$ and $cU^{[N+1]}$}
\label{app:globalphase}
Consider two propagators $U^{[N]}$ and $\hat{U}^{[N]}$ that differ only by a global phase factor $e^{{\rm{i}}\eta}$, i.e., $\hat{U}^{[N]}=e^{{\rm{i}}\eta}U^{[N]}$. In the construction of $cU^{[N+1]}$ and $c\hat{U}^{[N+1]}$, the phase factor manifests itself as an additional phase factor that is \textit{only} applied to the second block on the diagonal:
\begin{equation}
cU^{[N+1]} =\begin{pmatrix}
{\bf 1}^{[N]} & 0 \\
0 & U^{[N]}
\end{pmatrix}
\end{equation} 
but
\begin{equation}
c\hat{U}^{[N+1]} =\begin{pmatrix}
{\bf 1}^{[N]} & 0 \\
0 & e^{{\rm{i}}\eta}U^{[N]}
\end{pmatrix},
\end{equation} 
i.e. the {\it global} phase factor $e^{{\rm{i}}\eta}$ of a propagator $\hat{U}^{[N]}$ is transformed into a {\it relative} phase factor in the propagator 
$c\hat{U}^{[N+1]}$ of the augmented system and hence becomes experimentally measurable.

However, any {\it global} phase factor of $cU^{[N+1]}$ is  irrelevant in the considered experiments because it cancels when $cU^{[N+1]}$ is applied to the density operator [see Eq.~\eqref{rhoUsimp}]. Hence, such a global phase factor can be ignored when designing experimental pulse sequences to realize $cU^{[N+1]}$; see Table \ref{tab:c_prob}.

\subsection{Decomposition into individual droplet $\ell$ contributions}
\label{app:decmop}
\begin{figure}
\begin{center}
\includegraphics[width=1\linewidth]{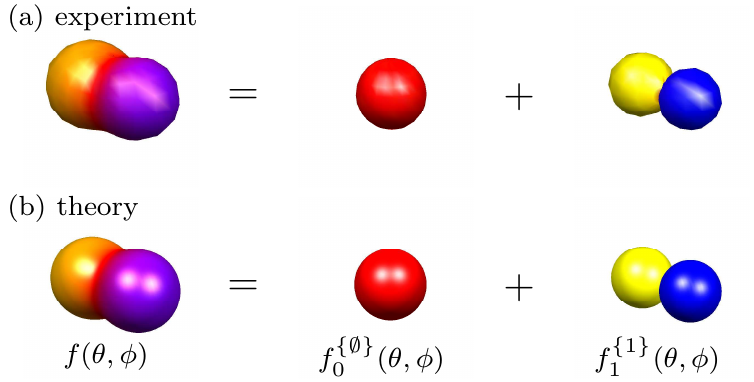}
\caption{(Color online) The Wigner representation $f(\theta, \phi)$ of a $[\tfrac{\pi}{2}]_x$ rotation propagator 
is decomposed into its 
contributions $f^{\{\emptyset\}}_0(\theta, \phi)$ and $f^{\{1\}}_1(\theta, \phi)$. \label{fig:decomp}} 
\end{center}
\end{figure}

\begin{figure}
\begin{center}
\includegraphics[width=1\linewidth]{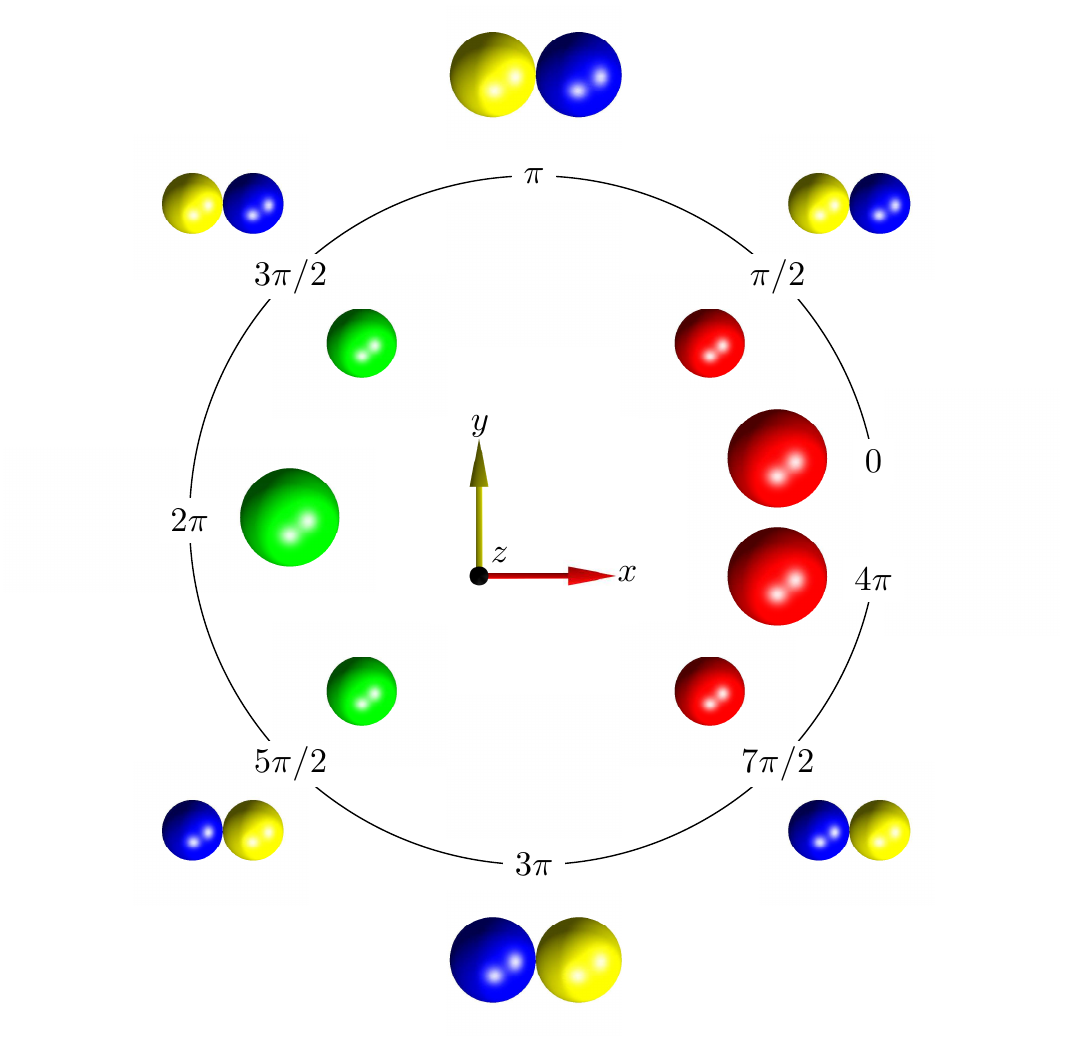}
\caption{(Color online) Theoretical Wigner functions $f(\theta,\phi)$ representing various propagators realizing  $x$ rotations decomposed into its 
contributions $f^{\{\emptyset\}}_0(\theta, \phi)$ (positioned along the inner circle)  and $f^{\{1\}}_1(\theta, \phi)$ (positioned along the outer circle).\label{fig:decomp_rot_sim}} 
\end{center}
\end{figure}

\begin{figure}
\begin{center}
\includegraphics[width=1\linewidth]{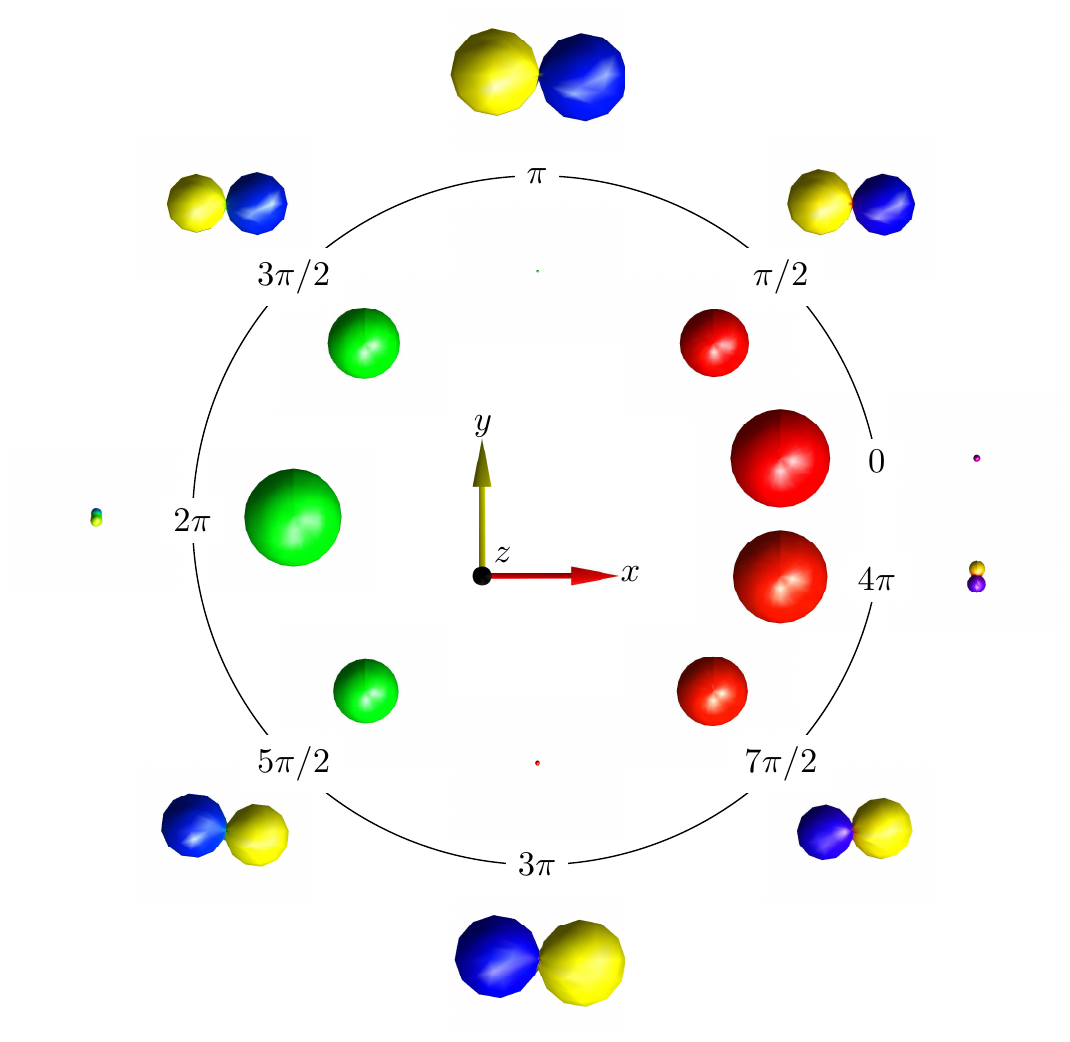}
\caption{(Color online) Experimentally reconstructed Wigner functions $f(\theta,\phi)$ representing various propagators realizing $x$ rotations decomposed into its
contributions $f^{\{\emptyset\}}_0(\theta, \phi)$ (positioned along the inner circle) and $f^{\{1\}}_1(\theta, \phi)$ (positioned along the outer circle). \label{fig:decomp_rot_exp}} 
\end{center}
\end{figure}

The proposed (experimental) Wigner process tomography scheme (see Results \ref{eq:prop_reconstr} 
and \ref{eq:prop_reconstr_NMR}) also allows one to decompose the function $f(\theta, \phi)$ 
into a part $f^{\{\emptyset\}}_0(\theta, \phi)$ corresponding to the identity operator
 (with rank $j=0$) and a part $f^{\{1\}}_1(\theta, \phi)$ corresponding to terms involving spin $I_1$  (with rank $j=1$). 
Although this creates some redundancy, the separation may help to analyze and to delineate different sources of errors in an experiment.
For example, the expected shape of $f^{\{\emptyset\}}_0(\theta, \phi)$ is a perfect sphere, and deviations of the experimentally reconstructed
shape of this droplet can give clues about the size of systematic and stochastic errors and  dominant experimental imperfections, such as pulse miscalibration, radio-frequency inhomogeneity or relaxation effects.
 Figure \ref{fig:decomp} illustrates the decomposition of $f(\theta, \phi)$ into $f^{\{\emptyset\}}_0(\theta, \phi)$ and $f^{\{1\}}_1(\theta, \phi)$
 for the case of a propagator corresponding to a $[\pi/2]_x$ rotation. Figures \ref{fig:decomp_rot_sim} and \ref{fig:decomp_rot_exp} show decomposed theoretical and experimental droplet functions $f^{\{\emptyset\}}_0(\theta, \phi)$ and $f^{\{1\}}_1(\theta, \phi)$  corresponding to the 
 combined droplet functions  $f(\theta, \phi)$ that are
shown in Fig.~\ref{fig:rotation} for the set of rotation angles 0, $\pi/2$, ..., $4\pi$.

\subsection{Explicit functional form of the droplets $f^{\{\emptyset \} }$ and $f^{\{1\}}$ for arbitrary rotations.
\label{app:functionalform}
}

For an arbitrary rotation with rotation angle $\Psi$ and a rotation axis defined by the normalized vector
\begin{equation}
\vec{n}=\left(\begin{array}{c}
n_x \\
n_y \\
n_z \end{array}\right),
\end{equation}
the propagator $U^{[1]}$ has the general form \cite{Kobzar2012,Goldman1988}  
\begin{equation}
U^{[1]}=\cos {\Psi\over 2} {\bf 1} - 2\ {\rm i} \ \sin {\Psi\over 2} (n_x I_x + n_y I_y + n_z I_z).
\end{equation}
Using the general relations 
$T_{00}=1/\sqrt{2}\ {\bf 1}$, 
$T_{1,-1}=I^-=I_x+{\rm i} I_y$, 
$T_{10}=\sqrt{2}\ I_z$, 
and $T_{11}=-I^+=-(I_x-{\rm i} I_y)$ 
between Cartesian spin operators and spherical tensor operators
\cite{EBW87,Garon15}
and the DROPS mapping between 
spherical tensor operators $T_{jm}$ and spherical harmonics $Y_{jm}$ (see Sec. \ref{chapt:summary_drops}), a straightforward calculation yields the following explicit functional forms for the 
$\ell=\{\emptyset \} $ and $\ell=\{1 \} $ 
components of the droplet function
\begin{equation}f(\theta, \phi)=f_0^{\{\emptyset \} }(\theta, \phi)+f_1^{\{1\}}(\theta, \phi):
\end{equation}
\begin{equation}
\label{eq:analy_f0}
f_0^{\{\emptyset \} }(\theta, \phi)=\sqrt{1\over{2 \pi}} \ \cos {\Psi\over 2}
\end{equation}
and
\begin{eqnarray}
f_1^{\{1\}}(\theta, \phi)=- {\rm i}\ \sqrt{3\over{2 \pi}} \ \sin {\Psi\over 2}\ \Big( n_x \sin \theta \cos\phi  \ \ \ \ \ \ \ \ \ \ \ \ \ \ \ \   \nonumber\\
+ \  n_y \sin \theta \sin\phi  + n_z \cos \theta    \Big). \ \ \ \ \ 
\end{eqnarray}
Note that the droplet function $f_0^{\{\emptyset \} }(\theta, \phi)$ is real-valued and independent of the polar and azimuthal angles $\theta$ and $\phi$. Hence it is represented either by a red (positive real) or green (negative real) sphere (dark gray and gray) that is centered at the origin in three-dimensional polar plots, see Figs. \ref{fig:decomp_rot_sim} - \ref{fig:error} and the color bar in Fig. \ref{fig:drops}.
The droplet function $f_1^{\{ 1 \} }(\theta, \phi)$ is purely imaginary and is represented in three-dimensional polar plots by a yellow (positive imaginary) and blue (negative imaginary)
sphere (light gray and black), which touch at the origin, see Figs. \ref{fig:decomp_rot_sim} - \ref{fig:error}. The vector connecting the centers of the yellow and blue spheres (light gray and black) is collinear with the rotation axis $\vec{n}$.
Due to the factor $\cos {\Psi\over 2}$, the function $f_0^{\{\emptyset \} }(\theta, \phi)$ is zero if the rotation angle $\Psi$ is an odd multiple of $\pi$.
Conversely, the function $f_1^{\{1 \} }(\theta, \phi)$ is zero if $\Psi$ is an even multiple of $\pi$ due to the 
 factor $\sin {\Psi\over 2}$; see Figs. \ref{fig:decomp}, \ref{fig:decomp_rot_sim}, and \ref{fig:error}.
 Note that some of the standard quantum gates are only identical to the propagator of a rotation up to a global phase factor. For example, the propagator of the Hadamard gate (see Table \ref{tab:c_prob}) is only up to e$^{-{\rm i}\pi/2}=-$i identical to a rotation with rotation angle $\Psi=\pi$ and rotation axis $\vec{n}$ with $n_x=n_z=1/\sqrt{2}$ and  $n_y=0$. This additional global phase factor changes the colors of the spheres of 
$f_1^{\{1 \} }(\theta, \phi)$ in Figs. \ref{fig:reconstr} and \ref{fig:1qubitgates}  from yellow and blue to  green (negative real) and red (positive real) (light gray and black to gray and dark gray), respectively.

\subsection{Periodicities of phase shift gate and rotation propagator for s spin 1/2}
\label{app:periodicity}
In  Secs.~\ref{results_phaseshift} and \ref{results_rot}, it was stated that for a spin 1/2 the propagator $U^{[1]}$ corresponding to a phase gate $U^{[1]}_{ph}(\gamma)$ and to a rotation $U^{[1]}_{rot}(\delta)$ has a periodicity of $2\pi$ for $\gamma$ but of $4\pi$ for $\delta$. This is a result of the fact that the matrix elements of 
\begin{equation}
U^{[1]}_{rot}(\delta)=\begin{pmatrix}
e^{-{\rm{i}} \delta/2} & 0 \\
0 & e^{{\rm{i}} \delta/2}
\end{pmatrix}
\end{equation}
are exponential functions of the half angle $\delta/2$ whereas the matrix elements of 
\begin{equation}
U^{[1]}_{ph}(\gamma) =\begin{pmatrix}
1 & 0 \\
0 & e^{{\rm{i}}\gamma}
\end{pmatrix}
\end{equation}
are exponential functions of the full angle $\gamma$, resulting in 
$U^{[1]}_{ph}(\gamma+2\pi)=U^{[1]}_{ph}(\gamma)$ but $U^{[1]}_{rot}(\delta+2\pi)=-U^{[1]}_{rot}(\delta)$ and $U^{[1]}_{rot}(\delta+4\pi)=U^{[1]}_{rot}(\delta)$.

\subsection{Analysis of propagator errors in the DROPS representation of propagators}
\label{app:errorterms}
\begin{figure}
\begin{center}
\includegraphics{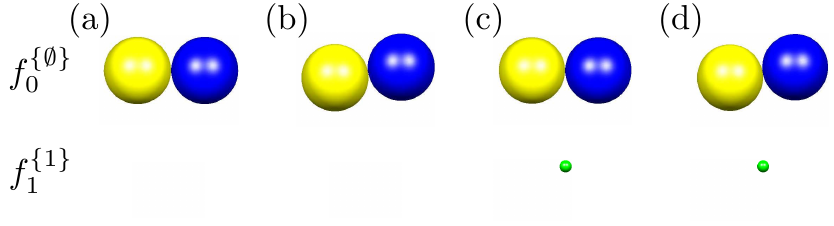}
\caption{(Color online) 
For the case of a system consisting only of a single ($N=1$) spin 1/2, the simulations show how different sources of propagator errors can be identified
in the DROPS representation: 
(a) Ideal case of a rotation with rotation angle $\Psi=\pi$ and rotation axis $\vec{n}$ with $n_x=1$ and $n_y=n_z=0$,
(b) rotation angle increased by 10\%,
(c) rotation axis deviating by an angle of $\pi/10$ from the $x$ axis resulting in $n_x=\cos (\pi/10)$ and $n_y= \sin (\pi/10)$ and  $n_z=0$,
and (d) simultaneous error of the flip angle as in case (b) and of the rotation axis as in case (c).
\label{fig:error}}
\end{center}
\end{figure}

Here, we consider possible advantages of the DROPS representation of propagators $U^{[N]}$ in the identification and quantification of propagator errors.
As demonstrated in this paper, the DROPS representation of a propagator can be measured directly by experimentally scanning
expectation values of given observables as a function of polar and azimuthal angles for a set of droplet functions $f^{(\ell)}$. (Alternatively, conventional process tomography schemes  \cite{Chuang1997,Poyatos1997,Schmiegelow2011,PhysRevA.64.012314,Altepeter2003,Gaikwad2018} could be used to
estimate the matrix elements of the propagator and to numerically calculate the corresponding DROPS representation of this matrix.)

As discussed in Sec. \ref{sec:discussion}, possible contributions to  errors of the experimentally sampled points of the droplet functions include
errors in the preparation of the initial density operator, in the implementation of the controlled version  $cU^{[N+1]}$ of the propagator $U^{[N]}$, noise in the detection process (see Fig. \ref{fig:exp}) and truncation effects in the automated integration and comparison of the spectra.
For simplicity, here we focus on an idealized scenario, where the errors of the tomography scheme are negligibly small and the scheme has been extended in analogy to
 \cite{Zhou2011} to the tomography of unknown propagators $U^{[N]}$ as discussed in Sec. \ref{sec:discussion}.
 Suppose we are trying to implement a desired target operator $U_{targ}^{[N]}$ by a new pulse sequence and we want to identify and quantify the 
 errors in the experimentally realized propagator $U_{exp}^{[N]}$.
The DROPS tomography of $U_{exp}^{[N]}$ provides a set of experimentally measured droplet functions, which (according to our assumption of negligible tomography errors) closely approach the ideal droplet functions corresponding to  
 $U_{exp}^{[N]}$.

If the system of interest consists of several spins, deviations of droplets from their ideal shapes corresponding to $U_{targ}^{[N]}$
provide information about specific error types. 
For example, if in a system with $N=2$ (see Fig. \ref{fig:drops}), errors of the propagator that only affect spin $I_1$ but not spin $I_2$ would be reflected by distortions of the droplet functions $f^{\{ 1\}}$ and $f^{\{ 12\}}$ but not of the droplets $f^{\{ 2\}}$.
For the simple case with $N=1$ and an ideal rotation with rotation angle $\Psi=\pi$ and rotation axis $\vec{n}$ with $n_x=1$ and $n_y=n_z=0$,
the effects of errors in $\Psi$ and/or $\vec{n}$ on the droplet functions are simulated in Fig. \ref{fig:error}.
Figure \ref{fig:error} (a) shows the ideal case, where the two spheres representing droplet $f^{\{ 1\}}$ are aligned with the $x$ axis and the droplet function 
$f^{\{\emptyset \}}$ is zero. Figure \ref{fig:error} (b) shows the case, where the actual rotation axis deviates by an angle $\pi/10$ in the $x$-$y$ axis (resulting in $n_x=\cos (\pi/10)$, $n_y= \sin (\pi/10)$, and  $n_z=0$), which is reflected
by a corresponding deviation of the $f^{\{ 1\}}$ orientation (the vector connecting the yellow and blue spheres [light gray and black]) from the $x$ axis.
In contrast, an error in the rotation angle is signaled by a non-vanishing droplet function $f^{\{\emptyset \}}$ (as well as a slightly reduced size of the $f^{\{ 1\}}$ droplet). This is shown in Fig. \ref{fig:error} (c) for the case of $\Psi=1.1 \pi$ and ideal rotation axis $n_x=1$ and $n_y=n_z=0$, where 
a small negative green (gray) sphere of $f^{\{\emptyset \}}$ indicates that the actual rotation angle $\Psi$ is larger than the desired rotation angle of $\pi$, see Eq. \eqref{eq:analy_f0}.
Finally, Fig. \ref{fig:error} (d) shows the effect of simultaneous errors of the rotation angle  and the rotation axis with $\Psi=1.1 \pi$, $n_x=\cos (\pi/10)$, $n_y= \sin (\pi/10)$, and  $n_z=0$, where the DROPS representation clearly shows both types of errors.
These examples show how the DROPS reconstruction can help to identify the source of systematic errors of a unitary propagator and to quantify the size of the respective errors. Random errors of the propagator can e.g. be identified by repeated measurements of droplet functions for the same polar and azimuthal angles.

\end{document}